\documentclass[12pt]{iopart}
\usepackage{epsfig}
\usepackage{graphicx}

\newcommand{\be     }{\begin{equation}}
\newcommand{\ee     }{\end{equation}}
\newcommand{\bd     }{\begin{displaymath}}
\newcommand{\ed     }{\end{displaymath}}

\newcommand{\bsigma }{{\mbox{\boldmath $\sigma$}}}

\newcommand{\btau   }{{\mbox{\boldmath $\tau$}}}
\newcommand{\bra    }{\langle}
\newcommand{\ket    }{\rangle}
\newcommand{\Bra    }{\left\langle}
\newcommand{\Ket    }{\right\rangle}

\newcommand{\ba     }{\ensuremath{\mathbf{a}}}

\newcommand{\bc     }{\ensuremath{\mathbf{c}}}
\newcommand{\bh     }{\ensuremath{\mathbf{h}}}

\newcommand{\bxi    }{{\mbox{\boldmath $\xi$}}}

\newcommand{\atanh  }{{\rm{atanh}}}

\newcommand{\betaslow}{\tilde{\beta}}

\begin{document}

\title[Slowly evolving random graphs]
{Slowly evolving random graphs II:
Adaptive geometry in finite-connectivity Hopfield models}
\author{B. Wemmenhove$^\P$ and N. S. Skantzos$^\S$ }
\address{\P ~Institute for Theoretical Physics, University of Amsterdam,
Valckenierstraat 65, 1018 XE Amsterdam, The Netherlands}
\address{\S ~ Departament de Fisica Fonamental, Facultat de Fisica, Universitat
de Barcelona, 08028 Barcelona, Spain}

\begin{abstract}
\noindent
We present an analytically solvable random graph model in which the connections between
the nodes can evolve in time, adiabatically slowly compared 
to the dynamics of the
nodes. We apply the formalism to finite connectivity attractor neural network
(Hopfield) models and we show that due to the minimisation of the frustration effects
the retrieval region of the phase diagram can be significantly enlarged. 
Moreover, the fraction of misaligned spins is reduced by this effect, and
is smaller than in the infinite connectivity regime. The main cause of
this difference is found to be the non-zero fraction of 
sites with vanishing local field when the connectivity is finite.
\end{abstract}

\section{Introduction}

Random graphs play an important role in a wide range of scientific
disciplines. The topological properties of graphs offer, in the
language of nodes and edges, a simple, yet powerful representation
of many different complex systems. For instance in biophysics,
simple models of random graphs have been used to predict the
function of a chemical component (a node or a `spin' on the graph)
given information about how different chemicals interact with one
another (graph connectivity) and their individual
functionality within the cell (spin orientation). 
In complex systems such as financial markets, one wishes to
understand the underlying laws and critical parameter
values relating to economic recession or success, given a certain
number of connections between traders and their individual
decisions. Examples of such complex systems are in abundance,
ranging from the aforementioned to ecological and linguistic
networks (for recent reviews on the subject see
\cite{dorogov,BarAlrev}). These systems share one common feature:
the sparse connectivity between the nodes. Every member of the
population is connected on average to only a small fraction of all other
members; yet a collective behaviour is observed.

Under this unifying framework, random graphs stand as an
interesting entity \emph{per se}. From a statistical mechanical
point of view such systems differ fundamentally from
fully-connected ones, especially in the parameter regime of finite
connectivity. In the latter case, even at the level of replica symmetry,
one finds that an order-parameter function is required to describe
equilibrium \cite{VB,kanter,MePa87,Wongsher}. 
Several applications of analytically solvable models
on finite connectivity random graphs can be found for
e.g.\@ error-correcting codes
\cite{KS,SvMSK}, cryptography \cite{KMS,cryptanalysis},
optimisation problems \cite{CM,MZ}, spin glasses \cite{MP,Ni04}
and neural networks
\cite{WC,IsaacNikos,scalefree}.

Physical systems however hardly ever maintain a static network
architecture: the structure of web-links changes continuously over
time, as does the connectivity between traders in stock markets, or
the proteinic interactions through the evolutionary process;
dynamically evolving graphs form the majority of complex systems in
nature. Studying such systems from an analytic viewpoint poses
an important theoretical challenge, since very
little is known about the dynamics of disordered systems with finite
connectivity. In this paper, we solve a simple model of a spin
system on a random graph in which the network architecture can
evolve in time, although on timescales adiabatically slow compared
to the dynamics of the spins. Thus spins are always at equilibrium
with respect to the dynamics of the connections. This important
class of models which can be solved exactly has been first studied
for fully-connected systems and Hopfield models
\cite{CoolenSherrington,PCS1,JBC,JABCP,CoolenUezu,FeldmanDotsenko,DotsFranzMezard},
whereas, curiously, the extension of these ideas to spin systems
evolving on a hierarchy of timescales reproduces the Parisi
full replica symmetry breaking scheme \cite{MourikCoolen}.

Here we
extend these ideas to finite-connectivity systems and, in
particular, we apply our formalism to Hopfield models.
We show that the retrieval region of the phase diagram becomes
significantly enlarged, and in fact, any finite degree of mobility of the graph can
lead to recall of a condensed pattern at any finite value of the 
storage ratio. To get a better understanding of 
this effect, we also compute the
fraction of misaligned spins and the fraction of sites with
vanishing local fields  
as functions of the ratio of the characteristic
temperatures
of the two dynamic processes. In the limit of extreme dilution our expressions
reproduce the results of \cite{WSC}.

\section{Model definitions}

The present model is an extension of the one presented in
\cite{WSC} where geometry was of an adaptive nature. There, one
could make important analytic simplifications owing to the choice
of `extreme dilution' scaling whereby each node in the graph was
connected on average to a vanishing fraction of other nodes
although this fraction still contained an infinite number of nodes.
Here, we extend the formalism of \cite{WSC} to consider the more
realistic (albeit analytically more involved) scenario in which for
every node there is a \emph{finite} average number of connections,
prescribed to a fixed number $c$. Thus, loops although rare are
still present.

To be precise, our model describes a graph of $i=1,\ldots,N$ nodes
associated with $N$ (fast) neuron variables
$\bsigma=(\sigma_1,\ldots,\sigma_N)$ with $\sigma_i
\in \{-1,1\}$. For every pair of neurons $(i,j)$ we consider the connectivity
variables $\bc=\{c_{ij}\}$ with $c_{ji} = c_{ij}$ and $c_{ii} = 0$,
which describe whether a connection between neurons $\sigma_i$ and
$\sigma_j$ is present ($c_{ij}=1$) or not ($c_{ij}=0$). We take
neuron variables to evolve according to a Glauber dynamics and at
equilibrium their energy is described by
\begin{equation}
H_{\rm f} (\bsigma,\bc)= -\sum_{i<j} \frac{c_{ij}}{c} \sum_{\mu = 1}^p \xi_i^\mu \xi_j^\mu
\sigma_i \sigma_j
\label{eq:fastH}
\end{equation}
The (quenched) variables $\{ \bxi^\mu \} \in \{-1,1\}^N$ with
$\mu=1, \ldots, p$ describe $p$ binary patterns which have been
imposed as attractors of the dynamics (stored Hopfield memories).
The value of the pattern interactions $\sum_\mu \xi_i^\mu\xi_j^\mu$
remains unchanged throughout the dynamics. Neuron variables
equilibrate with respect to this Hamiltonian at temperature
$1/\beta$ and are described by the partition function
\begin{equation}
Z_{\rm f}(\bc) = \sum_{\{\bsigma\}}{\rm e}^{-\beta H_{\rm f}(\bsigma,\bc)}
\label{eq:fastZ}
\end{equation}
Within the timescale in which neurons evolve to their equilibrium
state the connectivity variables $c_{ij}$ are effectively quenched.
On larger timescales though, these also evolve dynamically
(according to some Glauber prescription and preserving detailed
balance) with the total Hamiltonian
\begin{equation}
H_{\rm s}(\bc) =
-\frac{1}{\beta} \log Z_{\rm f}(\bc) + \frac{1}{\betaslow}
\log\left(\frac{N}{c}\right) \sum_{i<j} c_{ij}
\label{slowham}
\end{equation}
By this construction, energetically favourable configurations of
the connectivity matrix $\bc$ are taken to be those that minimise
the free energy of the neuron (fast) variables. The chemical
potential in (\ref{slowham}) ensures, as was shown in
\cite{WSC}, that on average there are $c$ connections per neuron.
Sparse connectivity then requires taking the limit $c/N\to 0$ while
our choice of finite connectivity scaling corresponds to
$c\sim\mathcal{O}(1)$.

The variables $\bc$, in turn, equilibrate at an inverse
temperature $\betaslow$ with a total partition function given by
\begin{equation}
Z_{\rm s} = \sum_{\{\bc\}} {\rm e}^{-\betaslow H_{\rm s}(\bc)} = 
\sum_{\{\bc\}}
\left[Z_{\rm f}(\bc)\right]^{\betaslow/\beta}
{\rm e}^{-\log(\frac{N}{c}) \sum_{i<j}c_{ij}}
\label{partsumslow}
\end{equation}
This partition sum
effectively contains $n=\betaslow/\beta$
replicated copies of the fast system, producing a replica theory
with nonvanishing but, generally, noninteger replica-dimension.
Hence, by definition, a large replica dimension $n$ corresponds to
low temperatures $\tilde{T}$
(or, equivalently, low-energies) of the slow system relative to that of the
fast. The formation of a particular graph configuration is then 
dominated by the Hamiltonian
(\ref{slowham}) rather than the thermal noise $\tilde{T}$.
The total construction of equations
(\ref{eq:fastH})-(\ref{partsumslow}) therefore rearranges the geometry
of the graph into optimised configurations $\bc$ such that the
retrieval performance of pattern recall can be potentially enhanced
(due to the minimisation of frustrated bonds).

At total equilibrium we will be
interested in the evaluation of the (slow) free energy
\begin{equation}
f=-\lim_{N\to\infty}\frac{1}{\tilde{\beta} N}\log Z_{\rm s}
\label{eq:f}
\end{equation}
which generates expressions for the system's macroscopic
observables.

\section{Calculation of the RS free energy}

To evaluate the partition function (\ref{partsumslow}),
first the trace over the connectivity variables is taken. This results in
\begin{equation}
Z_{\rm s}=\sum_{\bsigma^1\cdots\bsigma^n}\prod_{i<j}
\left[1+\frac{c}{N}{\rm e}^{\frac{\beta}{c}
\bxi_i\cdot\bxi_j\bsigma_i\cdot\bsigma_j}\right]
\end{equation}
where we introduced the notation
$\bsigma=(\sigma^1,\ldots,\sigma^n)$ with
$\bsigma\cdot\bsigma'=\sum_\alpha\sigma^\alpha\sigma'^\alpha$ and
$\bxi_i\cdot\bxi_j=\sum_{\mu=1}^p\xi_i^\mu\xi_j^\mu$ as usual. The
methodology required to proceed further from the above equation depends
on the scaling regime one considers. For systems with
extreme-dilution where $c\to\infty$ (while $c/N\to 0$)  \cite{WSC}
one can expand the above exponential and retain only the lowest two
moments, which for $N\to\infty$ describe the system's
thermodynamics fully. In contrast, for any finite $c$ moments of
order higher than 2 do not vanish. One is therefore required to
consider an order-parameter function.

At this stage, it turns out helpful to partition our graph into
$2^p$ sublattices \cite{WC,Hemmen}, $I_\bxi=\{i|\bxi_i=\bxi\}$ with
sublattice-averages $\bra \mathcal{F}(\bxi)\ket_{\bxi}=
\sum_{\bxi} p_{\bxi}\ \mathcal{F}(\bxi)$, where $p_\bxi=|I_{\bxi}|/N$.
Then, upon introducing into our equations the order-function
\begin{equation}
P_\bxi(\bsigma)=\frac{1}{|I_{\bxi}|}\sum_{i\in I_{\bxi}}\delta_{\bsigma,\bsigma_i}
\label{eq:Pdef}
\end{equation}
we obtain the following extremisation problem for (\ref{eq:f}):
\begin{eqnarray}
f&=&{\rm Extr}_{\{P_{\bxi}(\bsigma)\}}\ \left\{
-\frac{c}{2\tilde{\beta}}\Bra\!\!\Bra\sum_{\bsigma\bsigma'}
P_{\bxi}(\bsigma)P_{\bxi'}(\bsigma')\ {\rm e}^{\frac{\beta}{c}\bxi\cdot\bxi'
\bsigma\cdot\bsigma'}\Ket\!\!\Ket_{\bxi\bxi'}\right. \nonumber
\\
& &
\left.
+\frac{1}{\tilde{\beta}}\Bra\log\sum_{\bsigma}\exp\left[
c\Bra\sum_{\bsigma'}P_{\bxi'}(\bsigma')\ {\rm e}^{\frac{\beta}{c}\bxi\cdot\bxi'
\bsigma\cdot\bsigma'}\Ket_{\bxi'}\right]\Ket_{\bxi}\right\}
\label{eq:f_extr}
\end{eqnarray}
The order-function $P_{\bxi}(\bsigma)$ follows from the self-consistent
equation
\begin{equation}
P_{\bxi}(\bsigma)=\frac{\exp\left[c\bra\sum_{\bsigma'}
P_{\bxi'}(\bsigma')\ {\rm e}^{\frac{\beta}{c}\bxi\cdot\bxi'\bsigma\cdot\bsigma'}
\ket_{\bxi'}\right]}
{\bra\sum_{\bsigma''}\exp\left[c\bra\sum_{\bsigma'}
P_{\bxi'}(\bsigma')\ {\rm e}^{\frac{\beta}{c}\bxi^{''}\cdot\bxi'\bsigma^{''}\cdot\bsigma'}
\ket_{\bxi'}\right]\ket_{\bxi''}}
\label{eq:Pself}
\end{equation}

\subsection{Replica- and sublattice-symmetric assumptions}
\label{sec:P}

Our expressions (\ref{eq:Pself}), which describe the underlying
order-function of our system, depend on the unpleasant sublattice
index $\bxi$. Therefore solving this system of $2^p$ equations will
become increasingly hard with the number of patterns $p$. However,
if the system is in a state where there is a finite overlap only
with one pattern (say $\mu=1$), we expect that one can
make a `condensed ansatz', namely requiring the sublattice
distribution of replicated spins to depend only on the component of the
sublattice vector $\bxi$ corresponding to the condensed pattern,
i.e.,
\begin{equation}
P_{\bxi}(\bsigma) = P_{\xi_1}(\bsigma)
\label{condans}
\end{equation}
Taking the traces over the non-condensed pattern components
then reduces the order parameter equations (\ref{eq:Pself}) to
\footnote{from now on we will
use the proportionality symbol $\sim$ to express
distributions modulo their normalisation constant.}
\begin{eqnarray}
P_{\xi}(\bsigma) \sim
\exp \left\{ \frac{c}{2} \sum_{\btau}
[\cosh(\frac{\beta}{c} \bsigma \cdot \btau)]^{p-1}
\sum_{\xi'} P_{\xi'}(\btau) {\rm e}^{\frac{\beta}{c}\xi'\xi(\bsigma\cdot \btau)}
\right\}
\label{Peq_dummy}
\end{eqnarray}
By inspection of the RHS of (\ref{Peq_dummy}), one concludes that
the dependence on the remaining sublattice $\xi$ can only come in
in the form $P_{\xi}(\bsigma)=P(\xi \bsigma)$. Upon inserting this
form into the RHS of (\ref{Peq_dummy}), it follows that the
resulting equation is self consistent, as it should be:
\begin{eqnarray} P(\bsigma) =
\frac{\exp \left\{ c \sum_{\btau} P(\btau)\ {\rm e}^{\frac{\beta}{c} (\bsigma
\cdot \btau)}\ [\cosh(\frac{\beta}{c} \bsigma \cdot \btau)]^{p-1}
\right\}}
{\sum{\bsigma'}\exp \left\{ c \sum_{\btau} P(\btau)\ {\rm e}^{
\frac{\beta}{c}
(\bsigma'\cdot \btau)} [\cosh(\frac{\beta}{c} \bsigma' \cdot \btau)]^{p-1}
\right\}}
\label{eq:P}
\end{eqnarray}
This expression, devoid completely of pattern variables, allows for a
significant reduction of numerical costs.

We will now proceed further with assuming replica-symmetry (RS),
viz.\@ we require that permutation of spins within different
replica groups leaves the order function (\ref{eq:P}) invariant. In
general, this amounts to assuming that the order function acquires
e.g.\@ the form
\begin{equation}
P(\bsigma)=\int {\rm d} h\ W(h)\ \frac{{\rm e}^{\beta
h\sum_{\alpha}\sigma^{\alpha}}}{[2\cosh(\beta h)]^n}
\label{eq:RSgen}
\end{equation}
since the sum over replica groups acts in the desired
permutation-invariant way. In our case, where $n$ can take any
finite value, we find that for integer $n$ values,
significant simplifications occur if we rewrite replica-symmetry as
\begin{equation}
P(\bsigma) = \sum_{l=0}^n Q(2l-n)\ \delta_{(2l-n),\sum_\alpha \sigma^\alpha}
\label{eq:RSint}
\end{equation}
for some density $Q(x)$ which takes values only at a finite number of integer
points and with the normalisation constraint
$1=\sum_{l=0}^n {n \choose l}Q(2l-n)$.
The special case of (\ref{eq:RSint}) can also serve as a
test of the accuracy of the more general expression of (\ref{eq:RSgen}).

\subsection{Thermodynamic quantities for general $n$}

\begin{figure}[t]
\vspace*{40mm}
\begin{picture}(200,80)
\put(30,15){\includegraphics[height=5.7cm,width=6.2cm]{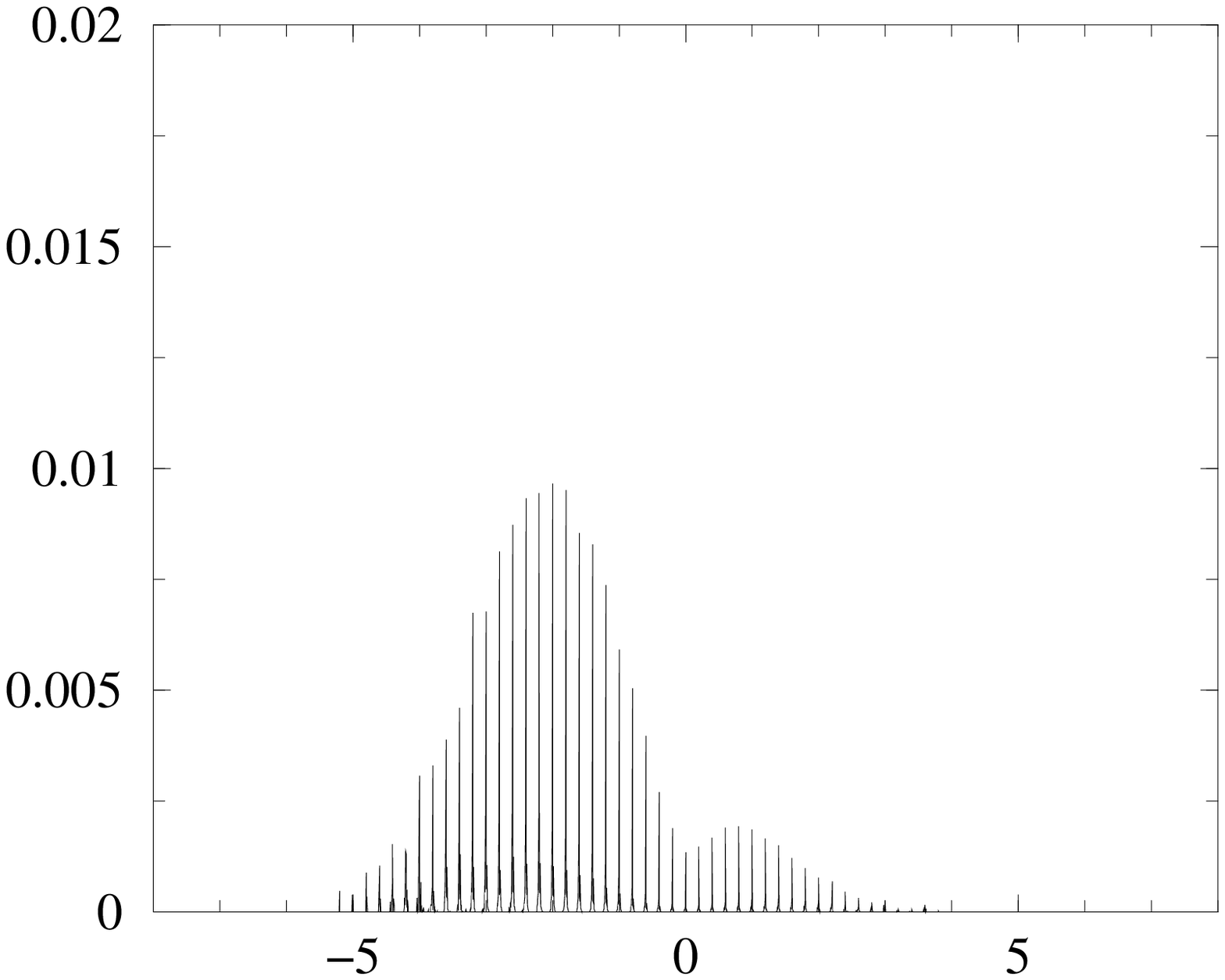}}
\put(245,15){\includegraphics[height=5.7cm,width=6.2cm]{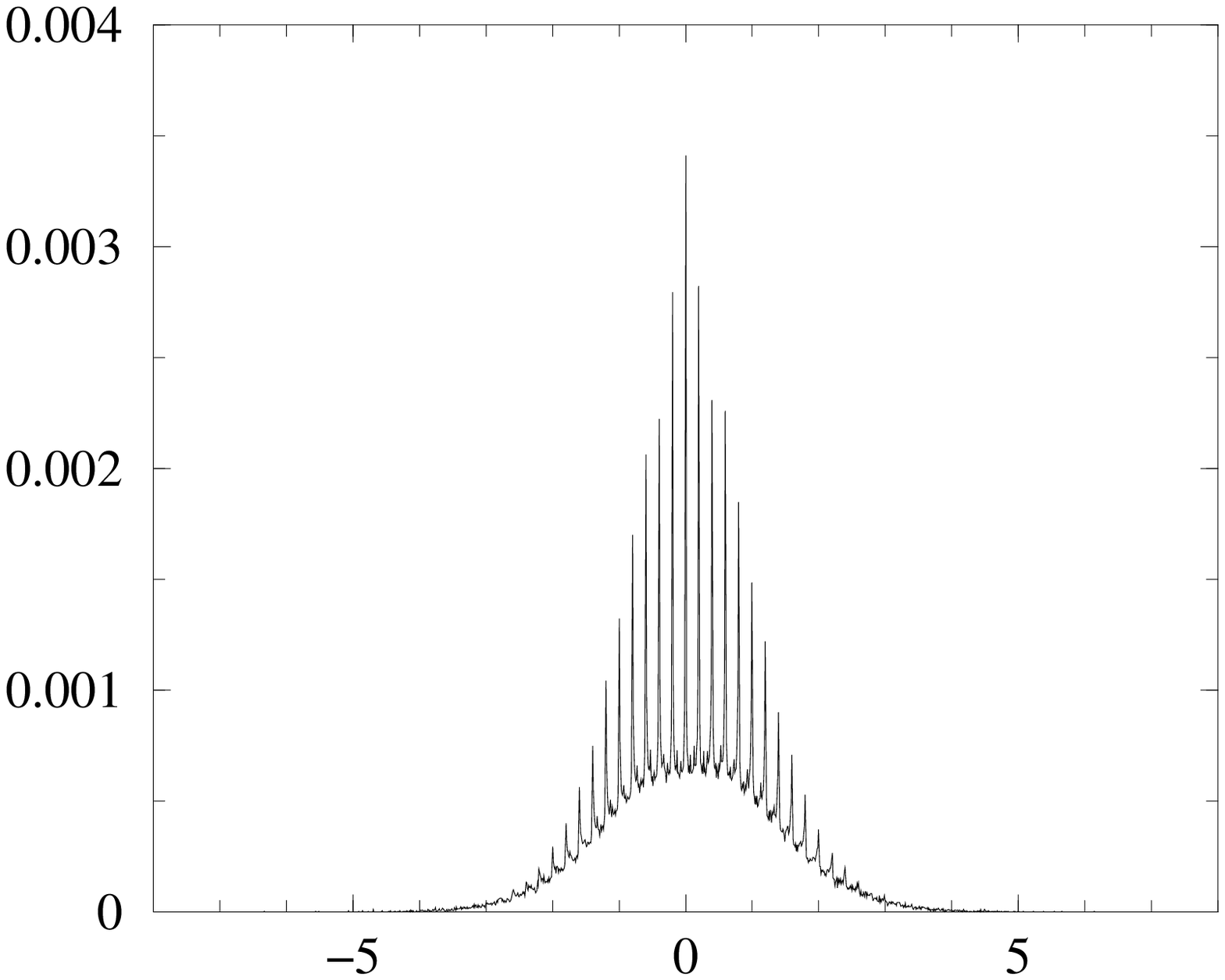}}
\put(125,6)  {\small $h$   }
\put(-2, 100){\small $W(h)$}
\put(330,6)  {\small $h$   }
\put(214,100){\small $W(h)$}
\end{picture}
\vspace*{-2mm}
\caption{The distribution $W(h)$ as obtained from equation (\ref{eq:W})
for $n=\frac{1}{10}$, $c=5$, $p=7$
and $T=0.05$ (left picture, corresponding to a ferromagnetic phase)
and  $T=0.2$ (right picture, corresponding to a spin-glass phase).}
\label{fig:W}
\end{figure}

Using the general-$n$ RS assumption (\ref{eq:RSgen}) we may convert
the self-consistent equation (\ref{eq:P}) into one for the
effective field distribution $W(h)$. To do so, first one expresses
(\ref{eq:f_extr}) in terms of $W(h)$, so that $f={\rm extr}_{\,
W(h)}\,f[W(h)]$ with:
\begin{eqnarray}
\hspace*{-20mm}
f[W(h)]=
-\frac{c}{2\tilde{\beta}}\int {\rm d}h{\rm d}h'\, W(h)W(h')\, \nonumber
\\
\times \Bra\cosh^{n}(\frac{\beta}{c}(p-2\nu))
\left[1+\tanh(\beta h)\tanh(\beta h')\tanh(\frac{\beta}{c}(p-2\nu))\right]^n\Ket_{\nu}
\nonumber
\\
+\frac{1}{\tilde{\beta}}\log\sum_{k=0}^\infty\frac{c^k {\rm e}^{-c}}{k!}
\Bra\cdots\Bra\ \left[\prod_l
\cosh^n(\frac{\beta}{c}(p-2\nu_l))\right] \right.\right. \nonumber
\\
\hspace*{-20mm}
\left.\left.\times\ \int [\prod_{l=1}^k  {\rm d}h_l\,W(h_l)]\left\{\sum_{\lambda=\pm 1}\left[
\prod_{l=1}^k\left(1+\lambda\tanh(\beta h_l)\tanh(\frac{\beta}{c}(p-2\nu_l))\right)
\right]\right\}^n\Ket_{\nu_1}\cdots\Ket_{\nu_k}
\nonumber
 \end{eqnarray}
with the averages
\begin{equation}
\Bra\mathcal{F}(\nu)\Ket_{\nu}=\sum_{\nu=0}^{p-1}
\left(\frac12\right)^{p-1} \left(\!\!\begin{array}{c} p-1 \\ \nu \end{array}\!\!\right)
\mathcal{F}(\nu)
\end{equation}
Variation now of the above with respect to $W(h)$ results in
\begin{eqnarray}
\lefteqn{W(h)\sim
\sum_{k=0}^\infty\frac{{\rm e}^{-c}c^k}{k!}
\Bra\cdots\Bra \ \left[\prod_{l=1}^k\cosh^n(\frac{\beta}{c}(p-2\nu_l))\right]
\right.\right.}\nonumber
\\
& &
\times\,\int [\prod_{l=1}^k {\rm d} h_l\,W(h_l)]\
\left[\sum_{\lambda=\pm 1}\prod_l\left(1+\lambda\tanh(\beta h_l)
\tanh(\frac{\beta}{c}(p-2\nu_l))\right)\right]^n  \nonumber
\\
& &
\left.\left.\times\
\delta\left[h-\frac1\beta\sum_{l}\atanh\left(\tanh(\beta h_l)\tanh(\frac{\beta}{c}
(p-2\nu_l))\right)\right]\Ket_{\nu_1}\cdots\Ket_{\nu_k}
\label{eq:W}
\end{eqnarray}
In the special limit $n\to 0$ this
expression recovers \cite{MePa87,kanter,WC} as it should.
Typical profiles of $W(h)$ are shown in figure \ref{fig:W}.

Using the ans\"atze (\ref{condans},\ref{eq:RSgen}) one
can also express the (condensed) retrieval and overlap order-parameters
in terms of $W(h)$:
\begin{equation}
m=\bra \xi\sum_{\bsigma}P_\xi(\bsigma)\,\sigma^1\ket_{\bxi}
=\int {\rm d} h\ W(h)\ \tanh(\beta h)
\label{eq:m}
\end{equation}
\begin{equation}
q=\bra \sum_{\bsigma}P_\xi(\bsigma)\,\sigma^1\sigma^2\ket_{\bxi}
=\int {\rm d} h\ W(h)\ \tanh^2(\beta h)
\label{eq:q}
\end{equation}
In figure \ref{fig:mq} we present numerical solutions of equations
(\ref{eq:m},\ref{eq:q}) (left panel).

From equations (\ref{eq:W},\ref{eq:m},\ref{eq:q}) we can identify
$W(h)=\delta(h)$ as the paramagnetic  (P) solution. Using
bifurcations analysis as in e.g.\@ \cite{WC} one can find possible
second-order transitions from this state. Thus, making an expansion
of (\ref{eq:W}) for small fields such that $\int
{\rm d} h\,W(h)\,h^\ell=\mathcal{O}(\epsilon^\ell)$ with $0<\epsilon\ll 1$
we find that transitions to ferromagnetic (F: $m,q\neq 0$) and
spin-glass regions (SG: $m=0$, $q\neq 0$) are respectively given by
\begin{eqnarray}
1=c\Bra \cosh^n(\frac{\beta}{c}(p-2\nu)\tanh(\frac{\beta}{c}(p-2\nu))\Ket_\nu
\label{eq:trans_m}
\\
1=c\Bra \cosh^n(\frac{\beta}{c}(p-2\nu)\tanh^2(\frac{\beta}{c}(p-2\nu))\Ket_\nu
\label{eq:trans_q}
\end{eqnarray}
In the limit $n\to 0$, when the evolution of connectivity variables
is dominated the thermal noise $\tilde{T}$, the microscopic
distribution of the $c_{ij}$ is uniform, and the calculation
is equivalent to a replica theory with quenched randomness. In this limit the
above transition lines then reduce indeed to the ones found in
\cite{WC}. However, one must also keep in mind here that for
increasing $n$, the phase diagram tends to be dominated by
first-order transitions
 \cite{sh}. In that case,
fields can no longer be regarded as small close to the transition
and a bifurcation analysis is not possible. One must then resort to
strictly numerical methods evaluating directly all observables of
interest.

\subsection{Thermodynamic quantities for integer $n$}

\begin{figure}[t]
\vspace*{40mm}
\begin{picture}(200,80)
\put(30,15){\includegraphics[height=5.4cm,width=5.8cm]{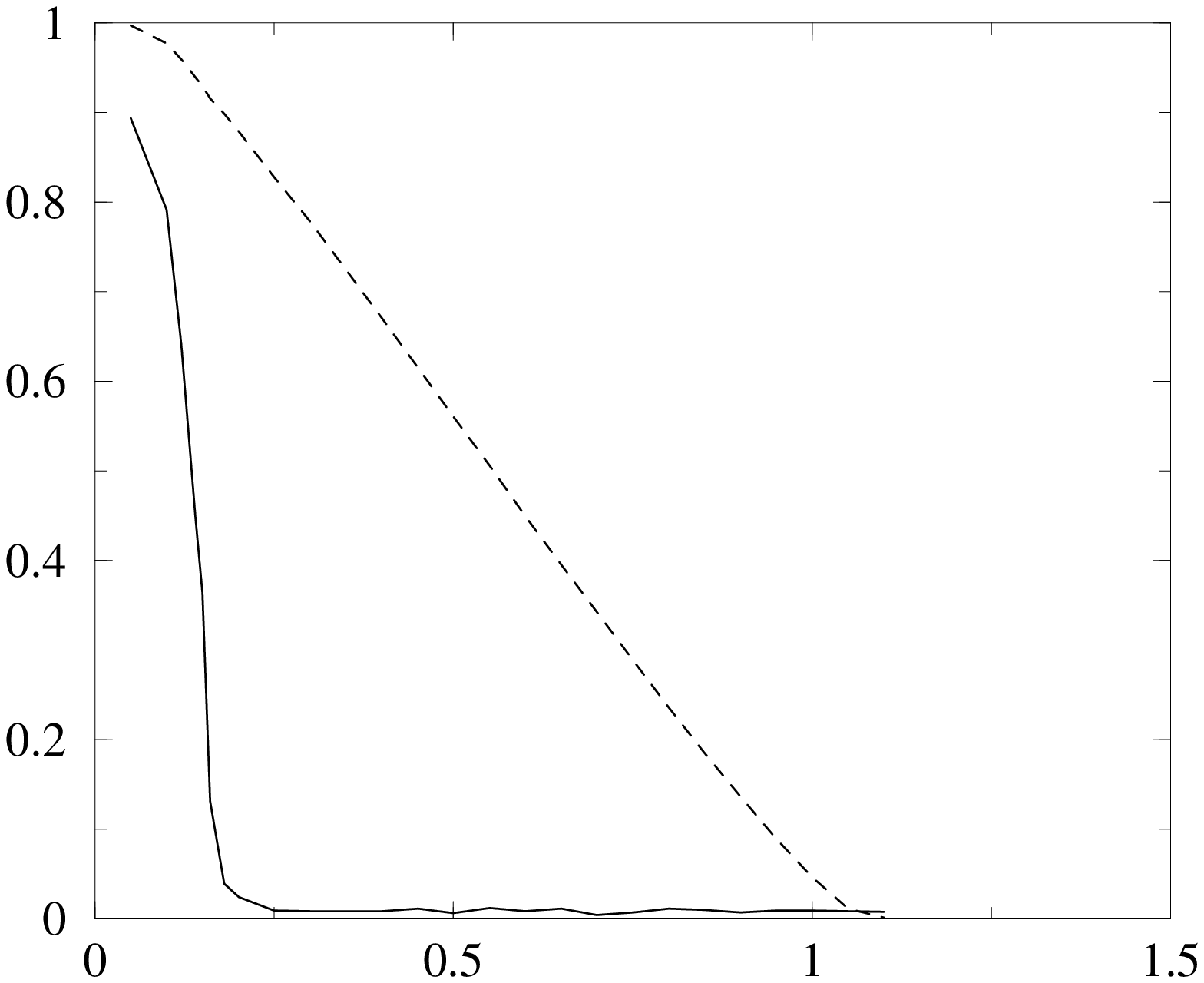}}
\put(245,15){\includegraphics[height=5.4cm,width=5.8cm]{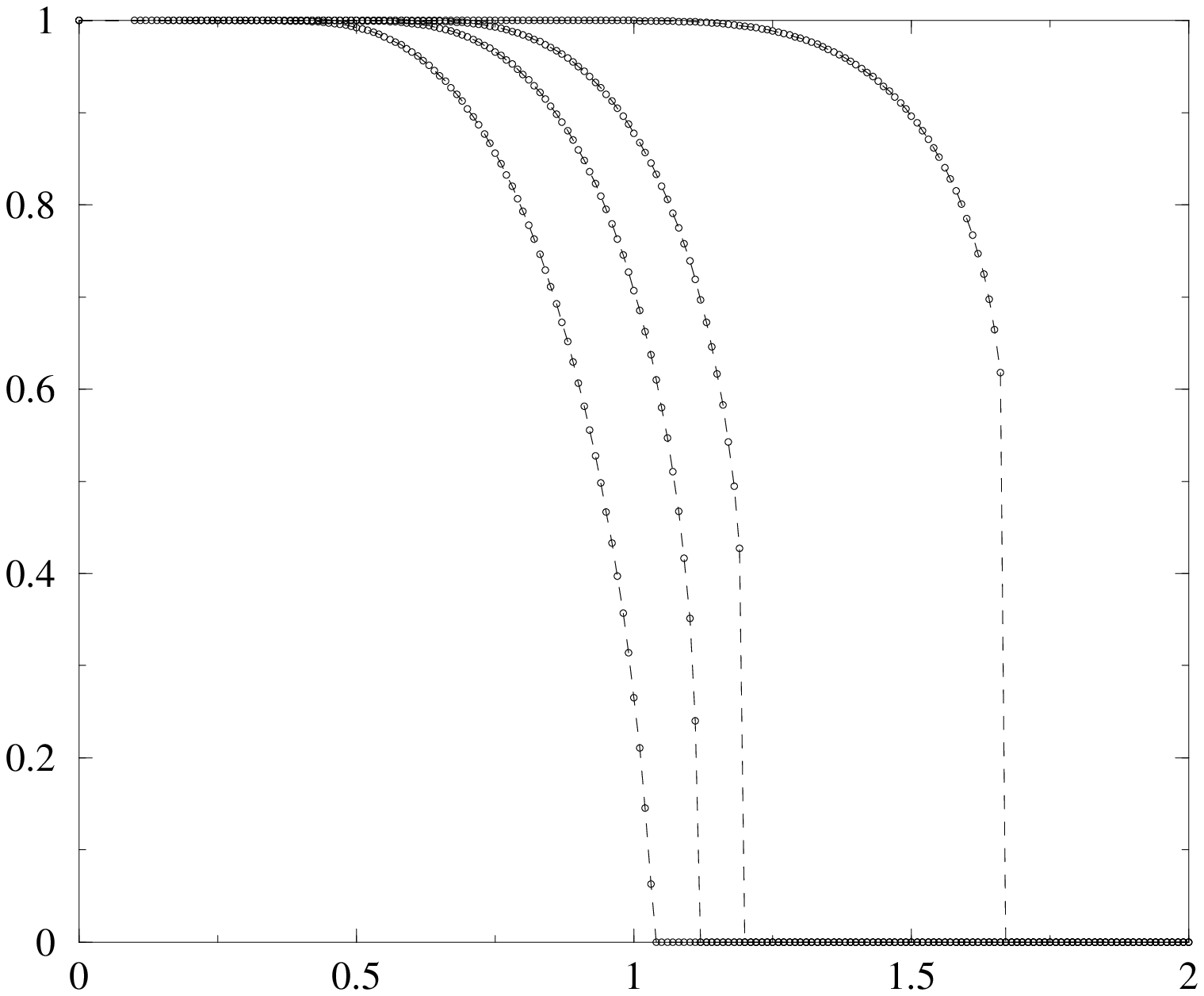}}
\put(115, 4){$T$}
\put(0, 90){$m,q$}
\put(325,4){$T$}
\put(225,90){$m$}
\end{picture}
\vspace*{-2mm}
\caption{First-order R$\to$SG transitions.
Left picture: The order parameters $m$ (solid) and $q$ (dashed)
for $n=0.1$ and $p/c=8/5$
using the general-$n$ expressions (\ref{eq:m},\ref{eq:q}).
Right picture: The order parameter $m$ using the integer-$n$ expressions
(\ref{eq:m_int}) with $n=2$, $c=5$ and $p=1,2,3,10$ (from left to right).}
\label{fig:mq}
\end{figure}

We will now consider the special case where $n$ is integer whereby
replica symmetry implies the order function $P(\bsigma)$ can only
have discrete arguments. Introducing the corresponding RS ansatz
(\ref{eq:RSint}) and replacing the Kronecker $\delta$-functions by
their integral representations allows us to rewrite the
self-consistent equation for $P(\bsigma)$ (\ref{eq:P}) in terms of
$Q(\ell)$, namely
\begin{equation}
Q(2s-n) \sim \exp \left\{ c \Bra  \sum_{l=0}^n Q(2l-n)
K(l, s,(p-2\nu))\Ket_{\nu} \right\}
\label{geq}
\end{equation}
with the constants (in a particular integral representation)
\begin{eqnarray}
\hspace*{-20mm}
K(l,s,(p-2\nu))= [2\cosh[\frac{\beta}{c}(p-2\nu)]]^n\, \sum_{j=0}^s
\sum_{k=0}^{n-s}\, { s \choose j}\, { n-s \choose k }
\tanh^{j+k}[\frac{\beta}{c}(p-2\nu)] \nonumber \\
\times \int_0^{2\pi}\frac{{\rm d}\omega}{2\pi} \cos^n(\omega)\,\tan^{j+k}(\omega)\,
\cos\left(\omega(2l-n) + \frac{\pi}{2}(3j+k)\right)
\end{eqnarray}
Equation (\ref{geq}) can be solved numerically by iteration. In
this representation, and also using the sublattice symmetric ansatz
(\ref{condans}), we find that the (condensed) retrieval overlap
order-parameter $m =  \sum_{\bsigma} P(\bsigma)
\sigma^1$ reads
\begin{equation}
m =
 \sum_{l=0}^n Q(2l-n) 2^n \int_{0}^{2\pi} \frac{{\rm d}\omega}{2\pi}
\sin[(2l-n)\omega] \tan(\omega)\cos^n(\omega)
\label{eq:m_int}
\end{equation}
whereas, the spin glass order parameter
$q = \sum_{\bsigma} P(\bsigma) \sigma^1 \sigma^2 $
is
\begin{equation}
q = \sum_{l=0}^n Q(2l-n) 2^n \int_{0}^{2\pi} \frac{{\rm d}\omega}{2\pi}
\cos[(2l-n)\omega]\tan^2(\omega)\cos^n(\omega)
\label{eq:q_int}
\end{equation}
Finaly, the free energy in this representation is
\begin{eqnarray}
f & = &-\frac{c}{2\tilde{\beta}} \Bra
\sum_{l,s=0}^n \left( \begin{array}{c} n \\ s
\end{array} \right)\, Q(2s-n)\, Q(2l-n)\,
K(l, s, (2p-\nu))\Ket_\nu \nonumber \\
&&+ \frac{1}{\tilde{\beta}}\, \sum_{l=0}^n \left( \begin{array}{c} n \\ l
\end{array} \right)\, Q(2l-n)\, \log Q(2l-n) 
\end{eqnarray}
Numerical solutions of equation (\ref{eq:m_int}) are shown in figure \ref{fig:mq}
(right panel).

\subsection{Phase diagrams}

\begin{figure}[t]
\vspace*{40mm}
\begin{picture}(200,80)
\put(100,15){\includegraphics[height=5.7cm,width=6.2cm]{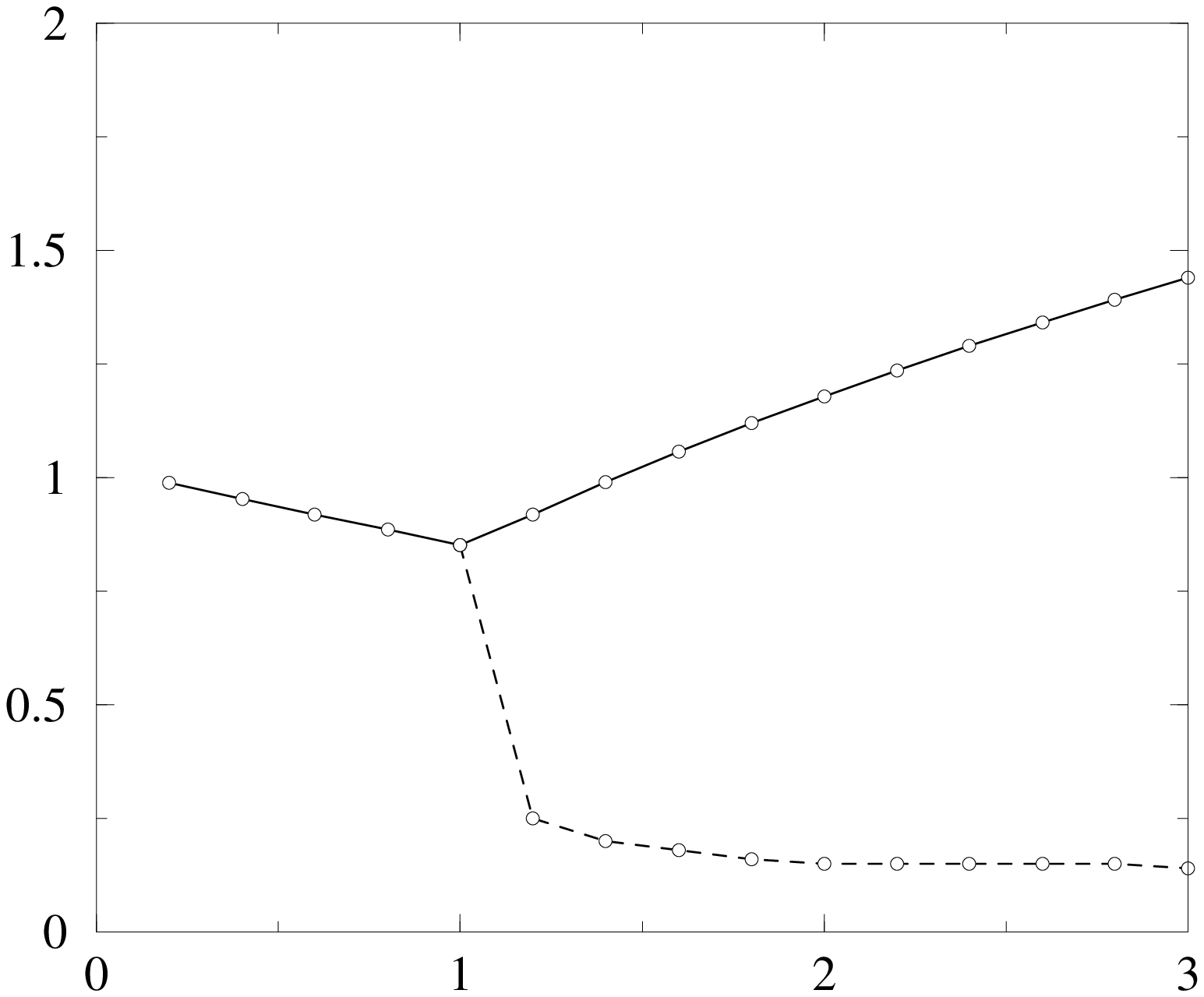}}
\put(190,6  ){$\alpha$}
\put(90,100){$T$}
\put(130,60 ){R}
\put(140,140){P}
\put(210,60){SG}
\end{picture}
\vspace*{-2mm}
\caption{Phase diagram for $n=0.1$ and $c=5$
in the space of $T$ and $\alpha=p/c$.
Solid and dashed lines represent second- and first-order transitions
respectively while only markers correspond to physical points (where
$p$ is integer). Even small values of $n$ lead to a significant enlargement of the
retrieval phase (c.f.\@ the phase diagram of \cite{WC} where $n\to 0$).
For sufficiently small values of $n$ the P$\to$R and P$\to$SG
transitions are second-order and the separating line is described
from (\ref{eq:trans_m},\ref{eq:trans_q}).}
\label{fig:phase_ngen}
\end{figure}

\begin{figure}[]
\vspace*{40mm}
\begin{picture}(200,80)
\put(30,15){\includegraphics[height=5.7cm,width=6.2cm]{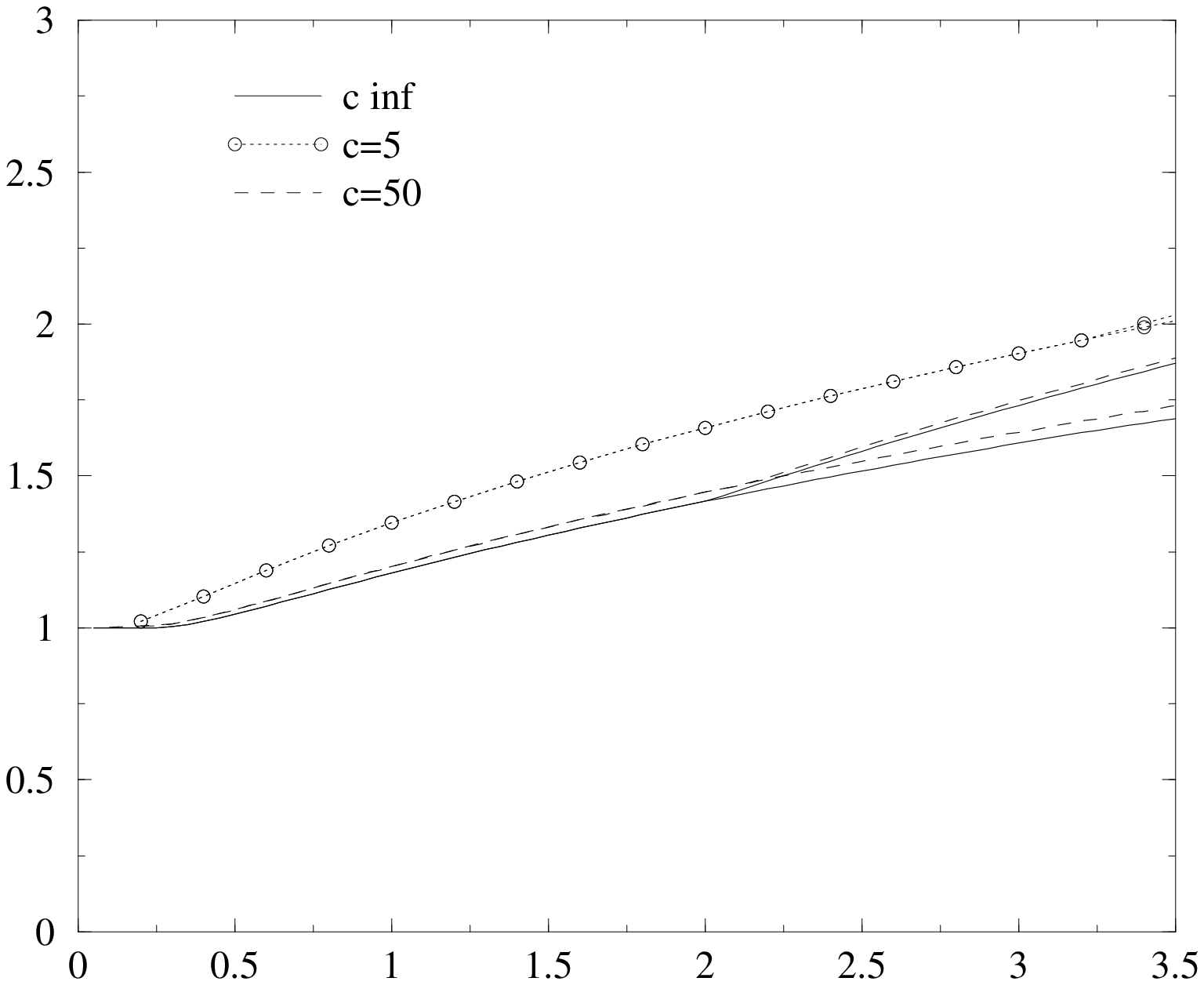}}
\put(245,15){\includegraphics[height=5.7cm,width=6.2cm]{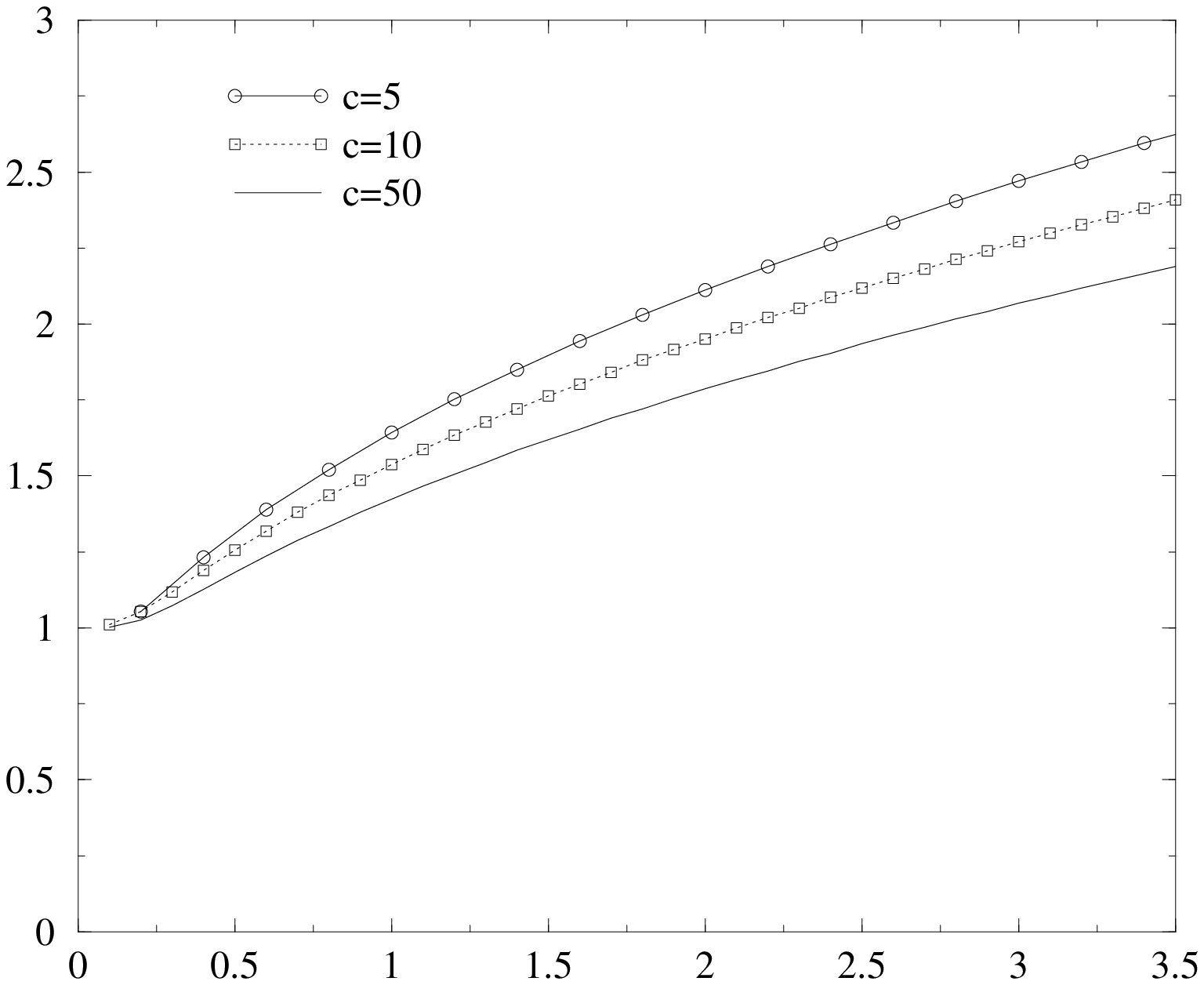}}
\put(120,6)  {$\alpha$}
\put(15, 100) {$T$}
\put(335,6)  {$\alpha$}
\put(230,100){$T$}
\put(60,120 ){\small  P}
\put(120,60 ){\small  R}
\put(280,120){\small  P}
\put(320,60 ){\small  R}
\put(190,110){\tiny  SG}
\end{picture}
\vspace*{-2mm}
\caption{Phase diagrams for $n=2$ (left) and $n=3$ (right)
in the space of $T$ and $\alpha=p/c$.
The transition lines have been derived by direct evaluation of the observables $m$
(\ref{eq:m_int}) and $q$ (\ref{eq:q_int}) and are of first order except for
very small $\alpha$.
(also see figure \ref{fig:mq}).
Increasing the value of the parameter $n$ leads to a shrinking of the SG phase.
This is an immediate result of the minimisation of frustration in the graph.}
\label{fig:phasediagrams_int}
\end{figure}

In figure \ref{fig:phase_ngen} we present the phase diagram of our
model for $n=0.1$. It is drawn in the $(\alpha,T)$ plane where
$\alpha=p/c$. Direct evaluation of the observables $m$ and $q$
(\ref{eq:W},\ref{eq:m},\ref{eq:q}) shows that the P$\to$R and
P$\to$SG transitions occur at the lines predicted by
(\ref{eq:trans_m}) and (\ref{eq:trans_q}). Generally, sufficiently
small values of $n$ lead to second-order P$\to$R and P$\to$SG
transitions, as was also the case for $c\to\infty$ \cite{WSC}. 
As expected, the transition R$\to$SG is first-order and
examples of $m$ along this transition are shown in figure
\ref{fig:mq}. Of special interest is the significant enlargament of
the R phase for any $n>0$. It physically implies that as soon as
the connectivity variables $\{c_{ij}\}$ can be `mobile' the graph
rearranges itself such that, for sufficiently low $T$,  pattern
recall can always be achieved. This effect is more apparent as the
ratio $n=\tilde{\beta}/\beta$ increases (see figures\@
\ref{fig:phasediagrams_int}): the range in the parameter space
where recall can be achieved increases, while the SG area shrinks.
Numerical evaluation of equations (\ref{eq:m_int},\ref{eq:q_int})
shows that large values of $n$ (i.e.\@ for $n>1$) the P$\to$R
transition line becomes first-order above some value of the storage capacity 
$\alpha$, which depends on $c$. In figure
\ref{fig:phasediagrams_int} we have plotted phase diagrams for
$n=2$ and $n=3$ for different values of the connectivity. Compared
to the extremely-diluted theory $c \to \infty$, we see that
finite-connectivity leads to larger regions in parameter space
where recall is possible (to illustrate this we have included the
corresponding $c\to\infty$ line in the left figure
where $n=2$). 
Physically, this can be understood on the basis of the 
number of connections per spin: for a system with
finite $c$, there are enough vacancies for the connectivity variables
to optimize the energetic contributions.
However, for $c\to\infty$, the number of vacancies for optimal 
configurations is less, creating an entropic countereffect.

\section{The fraction of misaligned spins}

Since our model aims to describe rearrangements of the connectivity
matrix $\{c_{ij}\}$ taylored to lead to optimal performance of
pattern recall, we expect on physical grounds, that this will be in
fact a result of a minimisation of the number of frustrated bonds
in the system. To measure this effect let us introduce the joint
distribution of spins and local fields
\begin{equation}
P_{\bxi}(\sigma,h)=\frac{1}{|I_{\bxi}|}\sum_{i\in I_{\bxi}}
\delta_{\sigma,\sigma_i}\ \delta[h-\sum_j\frac{c_{ij}}{c}\sum_{\mu}\xi_i^\mu\xi_j^\mu
\sigma_j]
\label{eq:joint_def}
\end{equation}
With (\ref{eq:joint_def}) we can now define the
fraction of misaligned spins
\begin{equation}
\phi=
\Bra \int_{-\infty}^0 {\rm d} h\ P_{\bxi}(1,h)+\int_0^{\infty}
{\rm d} h\ P_{\bxi}(-1,h)\Ket_{\bxi}
\label{eq:phi_def}
\end{equation}
and  the fraction of vanishing local fields
\begin{equation}
\psi=\Bra\sum_{\sigma\in\{-1,1\}} P_{\bxi}(\sigma,0)\Ket_{\bxi}
\label{eq:psi_def}
\end{equation}
The latter, in contrast with the $c \to \infty$ regime, is expected
not to vanish, since the distribution of local fields is a sum of delta peaks.
In the limit $c \to \infty$, this sum of delta peaks is traded for a 
continuous 
distribution, with a vanishing measure for the fraction of fields that are
exactly zero.
Expressions (\ref{eq:phi_def}) and (\ref{eq:psi_def}) in
combination will allow us to get an idea of the amount of frustration  in
our system.
Note however that a small 
fraction of misaligned spins does not necessarily imply a small amount of 
frustration: for a large number of patterns the interactions 
$\sum_\mu\xi^\mu_i \xi^\mu_j$
have a large absolute value and therefore with a high prbability spins 
prefer to align to their local field. 
On the contrary, a vanishing number of frustrated bonds does
imply a vanishing $\phi$. If variations in $n$ induce variations in $\phi$,
we expect this to be due to a change in the amount of frustration.

\subsection{The joint distribution of spins and local fields}

To evaluate the joint distribution of spins and local fields we
begin by inserting into the partition function (\ref{partsumslow})
appropriate delta functions to define the local fields in the
system
\begin{equation}
1 = \int \{ {\rm d}\bh_i \} \prod_i \delta[\bh_i - \sum_k \frac{c_{ik}}{c}
(\bxi_i \cdot \bxi_k) \bsigma_k]
\label{eq:localfields}
\end{equation}
with $\{{\rm d} \bh_i\}=\prod_{\alpha=1}^n {\rm d} h_i^\alpha$. The delta
function in the above expression introduces conjugate local fields
in our equations. Therefore, we are required to consider the more
general object
\begin{equation}
P_{\bxi}
(\bsigma,\bh,\hat{\bh})=
\frac{1}{|I_{\bxi}|}
\sum_{j\in I_{\bxi}} \delta_{\bsigma, \bsigma_j}\
\delta[\hat{\bh} - \hat{\bh}_j]\ \delta[\bh - \bh_j]
\label{eq:triplejoint}
\end{equation}
from which the joint distribution of spins and local fields
follows simply by integrating out the conjugate fields $\hat{\bh}$.
Given (\ref{eq:localfields}) and (\ref{eq:triplejoint}), we can now
take the trace over $\{c_{ij}\}$ in (\ref{partsumslow}) and arrive
at the free energy
\begin{eqnarray}
\fl
f={\rm Extr}_{\{P\}}\
\left\{-\frac{c}{2}\Bra\sum_{\bsigma\bsigma'}\int {\rm d} \bh 
{\rm d} \bh' {\rm d} \hat{\bh} {\rm d} \hat{\bh}'\
P_{\bxi}(\bsigma,\bh,\hat{\bh})\,P_{\bxi}(\bsigma,\bh,\hat{\bh})\
{\rm e}^{-\frac{i}{c}(\bsigma\cdot\hat{\bh}'+\bsigma'\cdot\hat{\bh})}\Ket_{\bxi\bxi'}\right. \nonumber
\\
\fl
\left.+\Bra\log\sum_{\bsigma}\int {\rm d} \bh {\rm d} \hat{\bh}\,
{\rm e}^{i\bh\cdot\hat{\bh}+\frac{\beta}{2}
\bh\cdot\bsigma+c\bra\sum_{\bsigma'}\int {\rm d}\bh'{\rm d} \hat{\bh}'\
P_{\bxi'}(\bsigma,\bh,\hat{\bh})\
{\rm e}^{-\frac{i}{c}(\bsigma\cdot\hat{\bh}'+\bsigma'\cdot\hat{\bh})}\ket_{\bxi'}}
\Ket_{\bxi}\right\}
\end{eqnarray}
where the extremisation problem leads to the self consistent equation for
$P_{\bxi}(\bsigma,\bh,\hat{\bh})$:
\begin{eqnarray}
\hspace*{-25mm}
P_{\bxi}(\bsigma,\bh,\hat{\bh}) \nonumber \\
\hspace*{-20mm}\sim
\exp\left\{ i\hat{\bh} \cdot \bh + \frac{\beta}{2} \bh \cdot \bsigma
         + c \left\bra \sum_{\bsigma'} \int {\rm d}\hat{\bh'} {\rm d}\bh'
         P_{\bxi'}(\hat{\bh'},\bh', \bsigma')
         {\rm e}^{-\frac{i}{c}(\bxi \cdot \bxi')[\hat{\bh}\cdot \bsigma'
         + \hat{\bh'} \cdot \bsigma]}\right\ket_{\bxi'} \right\}
\label{eq:self_Ptriple}
\end{eqnarray}
Let us now expand the exponential in the above right-hand side and integrate
over the conjugate fields $\hat{\bh}$:
\begin{eqnarray}
\hspace*{-25mm}
P_{\bxi}(\bsigma,\bh)\equiv
\int {\rm d}\hat{\bh}\,P_{\bxi}(\bsigma,\bh,\hat{\bh}) \nonumber \\
\hspace*{-10mm} \sim
{\rm e}^{\frac{\beta}{c}\bh\cdot\bsigma}\sum_{k\geq 0}\frac{{\rm e}^{-c}
c^k}{k!}
\Bra \cdots \Bra
\sum_{\btau_1\cdots\btau_k}\int [\prod_{l=1}^k {\rm d}\bh_l {\rm d}
\hat{\bh}_l
P_{\bxi'_l}(\btau_l,\bh_l,\hat{\bh}_l)]\,{\rm e}^{-\frac{i}{c}\sum_l\bxi\cdot\bxi'_l
\hat{\bh}'_l\cdot\btau_l} \right. \right. \nonumber \\
\left.
\left. \times \delta\left[\bh-\sum_{l=1}^k\frac{\bxi\cdot\bxi'_l}{c}
\btau_l\right]\Ket_{\bxi_1}\cdots \Ket_{\bxi_k} 
\end{eqnarray}
Using now the identity $\int {\rm d}\hat{\bh} {\rm d}\bh {\rm e}^{-i\ba \cdot \hat{\bh}}
P_{\bxi}(\bsigma,\bh,\hat{\bh})
= {\rm e}^{\frac{\beta}{2} \ba \cdot \bsigma} \int {\rm d}
\hat{\bh} {\rm d}\bh\ P_{\bxi}
(\bsigma,\bh,\hat{\bh})$
we can write
\begin{eqnarray}
\fl
P_{\bxi}(\bsigma,\bh) \sim {\rm e}^{\frac{\beta}{2} \bh \cdot \bsigma}
\sum_{k=0}^\infty \frac{c^k {\rm e}^{-c}}{k!} \Bra\cdots\Bra
\left[ \sum_{\btau_1\cdots\btau_k} \prod_{l=1}^kP_{\bxi_l'}(\btau_l) \right]
\exp\left\{\frac{\beta}{2c} \sum_{l=1}^k  (\bxi \cdot \bxi_l')
\bsigma \cdot \btau_l\right\}\right.\right. \nonumber
\\
\left.\left.
\times\
\delta \left[ \bh - \frac{1}{c}\sum_{l=1}^k (\bxi \cdot \bxi_l') \btau_l
\right] \Ket_{\bxi_1}\cdots\Ket_{\bxi_k}
\label{eq:joint_n}
\end{eqnarray}
where $P_{\bxi}(\bsigma)$ is the order function we defined in (\ref{eq:Pdef}).
Equation (\ref{eq:joint_n}) represents the joint distribution of $n$ spins and local
fields. From here, we may find the single-replica joint distribution
(\ref{eq:joint_def}) by integrating out $n-1$ replicas. The result is
\begin{eqnarray}
\hspace*{-25mm}
P_{\bxi}(\sigma,h) \sim
\sum_{k=0}^\infty \frac{c^k {\rm e}^{-c}}{k!} \left\bra \ldots \left\bra
\left[ \sum_{\btau_1\cdots\btau_k} \prod_{l=1}^k P_{\bxi_l'}(\btau_l) \right]
\prod_{\alpha=2}^{n} \left\{ 2\cosh\left[ \frac{\beta}{c}
\sum_{l=1}^k (\bxi \cdot \bxi_l')\tau_l^\alpha\right] \right\}
\right. \right. \nonumber \\
\left. \left.
 \times
\exp \left\{ \frac{\beta}{c} \sum_{l=1}^k  (\bxi \cdot \bxi_l')
\sigma \tau_l^1 \right\}
\delta \left[ h - \frac{1}{c}\sum_{l=1}^k (\bxi \cdot \bxi_l') \tau_l^1
\right] \right\ket_{\bxi_1'} \ldots \right\ket_{\bxi_k'}
\label{jodist}
\end{eqnarray}
This expression depends only on the order-function
$P_{\bxi}(\bsigma)$. To proceed further one is now required to
substitute either the general-$n$ replica- and sublattice-symmetric
assumptions (\ref{eq:RSgen},\ref{eq:W}), or, the corresponding
integer-$n$ ones (\ref{eq:RSint},\ref{geq}). For general $n$ we
find that tracing over the spins $\{\btau_l\}$ results in lengthy
expressions of questionable practical value.
For integer $n$ however, spins need not be necessarily traced: the
fact that their dimensionality is prescribed to the well-defined
integer value $n$ allows us to proceed further and rewrite
(\ref{jodist}) in a numerically tractable form. Using the ansatz
(\ref{condans}) in the RHS of (\ref{jodist}) and the RS expression
(\ref{eq:RSint}) we obtain
\begin{eqnarray}
\fl
P_{\xi}(\sigma,h)\sim 
\sum_{k=0}^\infty \frac{{\rm e}^{-c}c^k}{k!} \Bra\Bra\Bra\
\prod_{\alpha=2}^n\left\{2\cosh\left[\frac{\beta}{c}\sum_{l=1}^k
(p-1-2\nu_l+\xi\xi'_l)\tau^\alpha_l\right]\right\}\ \right.\right.\right.
\nonumber
\\
\fl
\left.\left.\left.
\times\
{\rm e}^{\frac{\beta}{c}\sigma\sum_l[p-1-2\nu_l+\xi\xi']\tau^1_l}\
\delta\left[h-\frac1c\sum_{l=1}^k[p-1-2\nu_l+\xi\xi'_l]\tau^1_l\right]
\Ket_{\nu_1\cdots\nu_k}
\Ket_{\btau^1\cdots\btau^k;\xi'_1\cdots\xi'_k}\Ket_{\xi'_1\cdots\xi'_k}
\label{eq:joint_final_gen}
\end{eqnarray}
with the abbreviated averages
\begin{equation}
\Bra \mathcal{F}(\nu)\Ket_{\nu}=
\sum_{\nu=0}^{p-1}\left(\frac12\right)^{p-1}
\left(\!\!\begin{array}{c} p-1 \\ \nu \end{array}\!\!\right)\
\mathcal{F}(\nu)
\label{eq:distr_nu}
\end{equation}
\begin{equation}
\Bra \mathcal{F}(\btau)\Ket_{\btau;\xi}=
\sum_{\btau}\sum_{s=0}^nQ(2s-n)\ \delta_{\sum_{\alpha}\xi\tau^\alpha,2s-n}\
\mathcal{F}(\btau)
\label{eq:distr_btau}
\end{equation}
\begin{equation}
\Bra \mathcal{F}(\xi)\Ket_{\xi}=\sum_{\xi}
[\frac12\,\delta_{\xi,1}+\frac12\,\delta_{\xi,-1}]\ \mathcal{F}(\xi)
\label{eq:distr_xi}
\end{equation}
Once the density $Q(\ell)$ has been obtained from (\ref{geq}) one
can evaluate the distribution (\ref{eq:joint_final_gen}) in the
spirit of population dynamics \cite{MP}: one considers a population
of triplets $\{\nu,\btau,\xi\}$. One then selects a number $k$ from
a Poisson distribution of mean $c$ and chooses $l=1,\ldots,k$
triplets $\{\nu_l,\btau_l,\xi'_l\}$ each according to the
distributions
(\ref{eq:distr_nu},\ref{eq:distr_xi},\ref{eq:distr_btau}). One then
evaluates the location and weight of the local field $h$ and thus
estimates the probability $P_{\xi}(\sigma,h)$ (with $\sigma$ and
$\xi$ fixed and modulo the normalisation coefficient). This
process is repeated until convergence.

\subsection{Numerical results}

\begin{figure}[t]
\begin{picture}(80,160)
\put(10,15){\includegraphics[height=5.0cm,width=4.5cm]{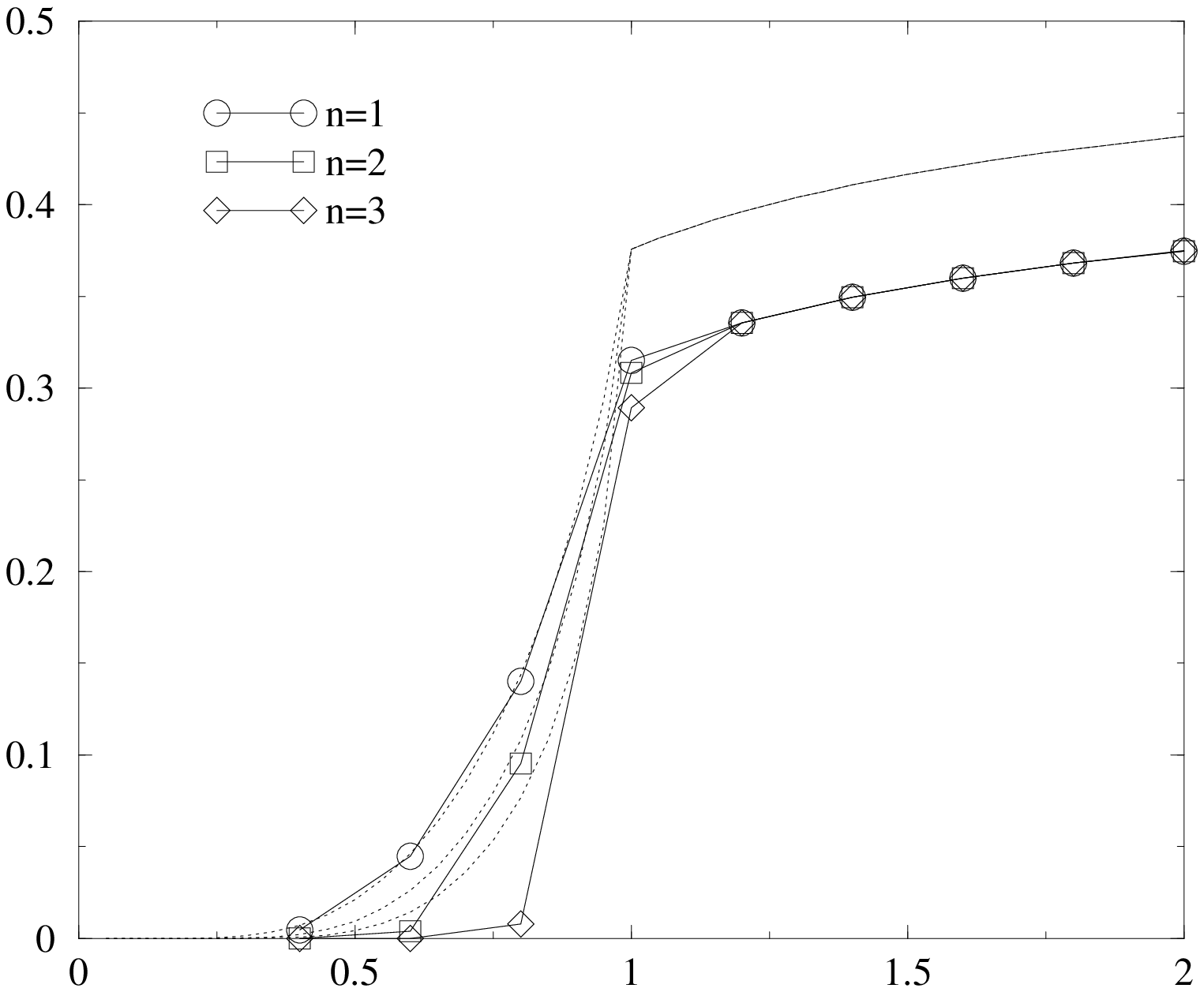}}
\put(160,15){\includegraphics[height=5.0cm,width=4.5cm]{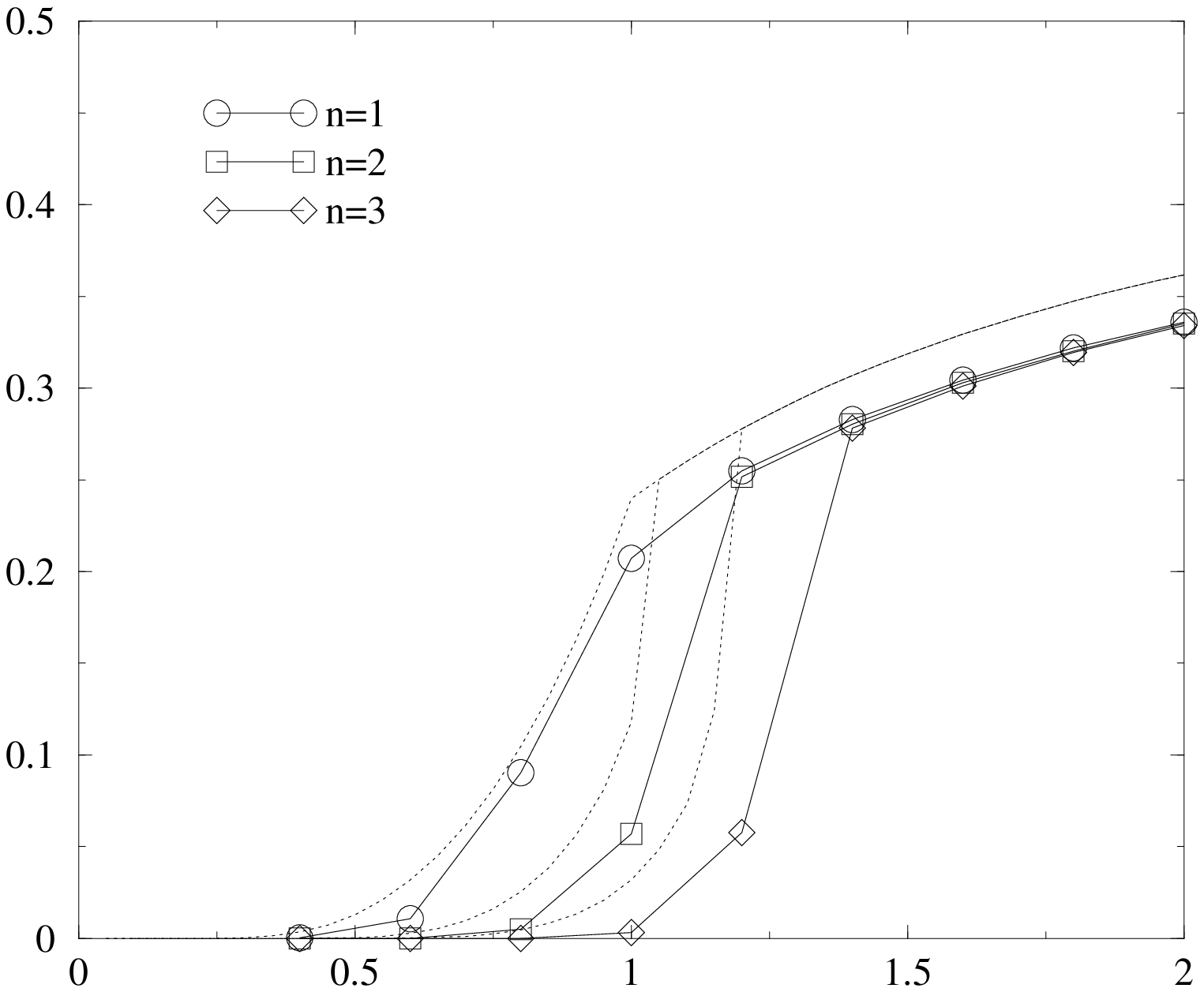}}
\put(310,15){\includegraphics[height=5.0cm,width=4.5cm]{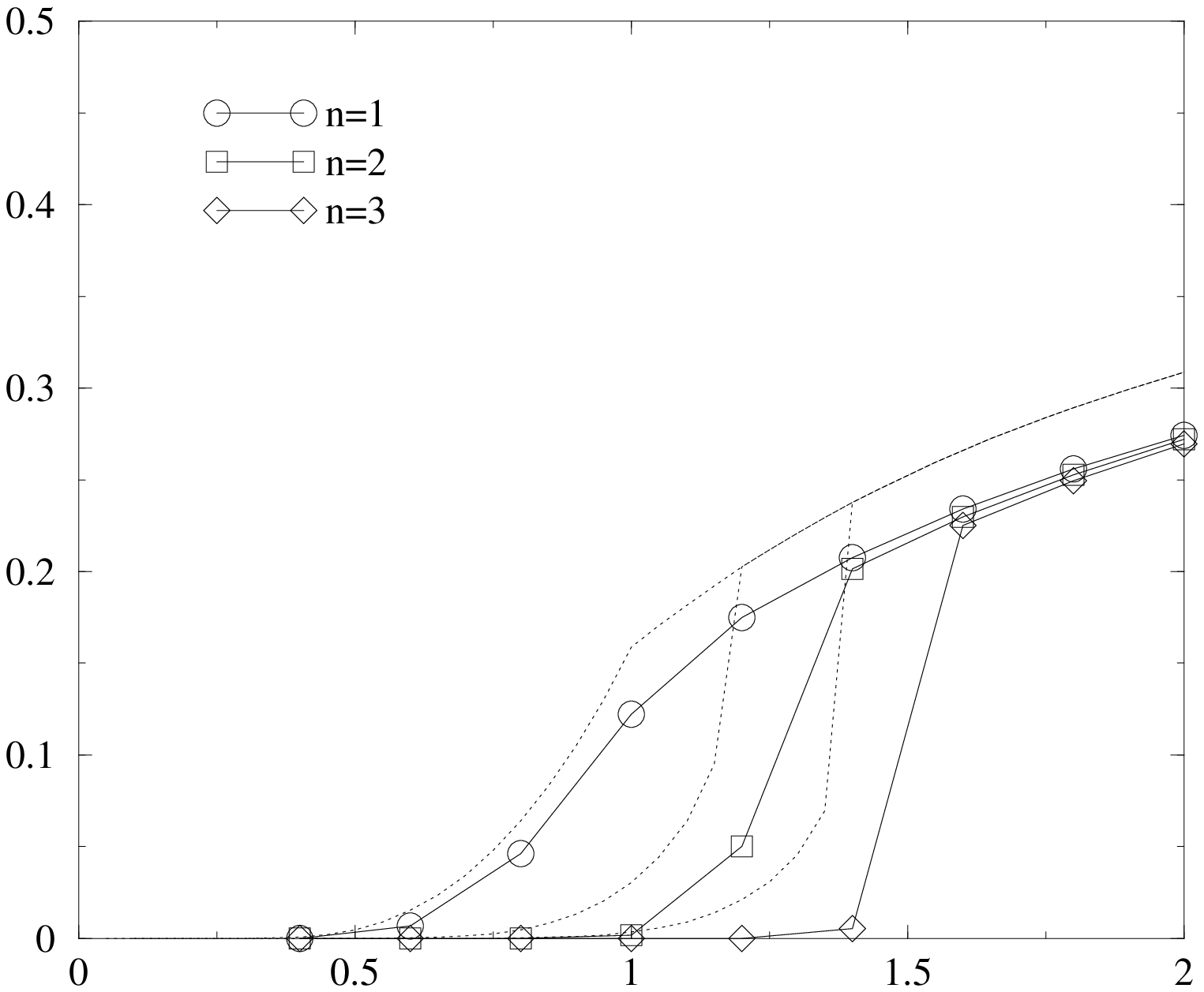}}
\put(70,0){{\small $T$}}
\put(0,90){{\small $\phi$}}
\put(220,0){{\small $T$}}
\put(150,90){{\small $\phi$}}
\put(370,0){{\small $T$}}
\put(300,90){{\small $\phi$}}
\end{picture}
\caption{Fraction of misaligned spins $\phi$ (\ref{eq:phi_def}) for
$\alpha=0.1,0.5,1$ (from left to right). In all figures $c=10$
and results are compared to the extreme dilution limit $c \to
\infty$ (dotted lines), for $n=1,2,3$.}
\label{frac1}
\end{figure}

In figure \ref{frac1} we plot the fraction of misaligned spins
(\ref{eq:phi_def}) as a function of the (spin) temperature $T$ for $c=10$
and $\alpha=0.1$, $0.5$ and $1$ respectively. As expected on
physical grounds, we see that larger values of $n$, i.e.\@ lower
temperatures $\tilde{T}$ (or, lower timescales, or lower energies
equivalently) associated to the slow wiring variables $c_{ij}$,
alignment improves. This effect is due to the ordering of the
$c_{ij}$, reducing the amount of frustration.
In all cases we compare the results
with those for $c\to\infty$ (as derived in \cite{WSC}). 
Clearly, also for finite fixed $c$, $\phi$ decreases as a function of 
$\alpha$, which is
in fact a result of the scaling choice of the interactions 
$\sum_{\mu} \xi^\mu_i \xi^\mu_j/c$. The
expectation of their absolute value increases with $p$, as do the local fields,
resulting in a high probability of alignment. 
We find
that finite connectivity values of $\phi$ are always upper-bounded by
those of $c\to\infty$. For temperatures $T$ sufficiently
large, such that the system is in a paramagnetic state, the noisy
dynamics of the spins dominates on the evolution of the graph and
$\phi$ becomes independent of the ratio $n$.

\begin{figure}[t]
\begin{picture}(80,190)
\put(120,10){\includegraphics[height=6.0cm, width=7.0cm]{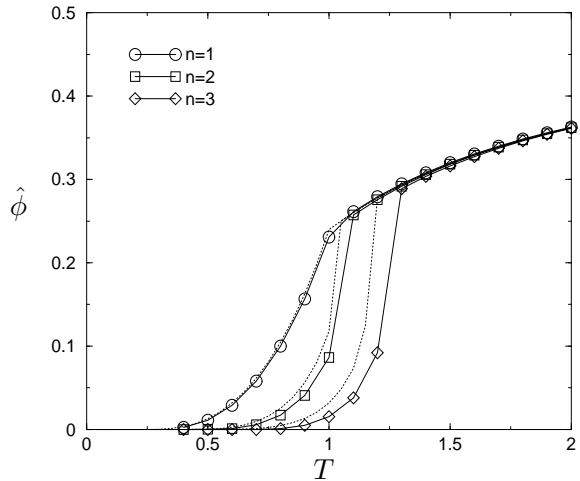}}
\put(220,0){\small $T$}
\put(105,100){\small $\hat{\phi}$}
\end{picture}
\caption{Fraction of misaligned spins plus half of the fraction of
spins with zero local field (\ref{eq:phihat_def}) for $c=14$,
$\alpha=0.5$, $n=1,2,3$, as compared to the case $c \to \infty$
(dotted lines). Differences between the two graphs are confined to temperature values
close to the phase boundary.}
\label{frac2}
\end{figure}

The difference between finite and infinite connectivity, most
clearly observed in the paramagnetic regime (large $T$), is found
to be mainly due to a nonvanishing fraction $\psi$ of sites that
have a zero local field (\ref{eq:psi_def}). This means that for 
such sites the energetic cost of either of the two possible 
orientations is identical. To understand the origin of the 
difference between finite-$c$ and $c\to\infty$ let us consider the quantity
\begin{equation}
\hat{\phi}= \phi + \frac12 \psi
\label{eq:phihat_def}
\end{equation}
which represents the fraction of frustrated sites to which half of 
the `indeterminate' spins have been added. 
This quantity is plotted in figure \ref{frac2} for $c=14$ and $\alpha=0.5$,
and is compared to $\phi$ in the limit of $c \to \infty$ calculated
in \cite{WSC}. We see that the finite- and infinite-$c$ plots become
almost identical apart from a small difference close to the
R$\to$P transition. This hints that, as far as the fraction of misaligned 
spins is concerned, $\psi$ is the main discriminator 
between the 
two scaling regimes and that the choice of obtaining the $c\to\infty$ fraction $\phi$ via
(\ref{eq:phihat_def}) is quite accurate.

\begin{figure}[t]
\begin{picture}(80,160)
\put(15,15){\includegraphics[height=5.0cm,width=4.5cm]{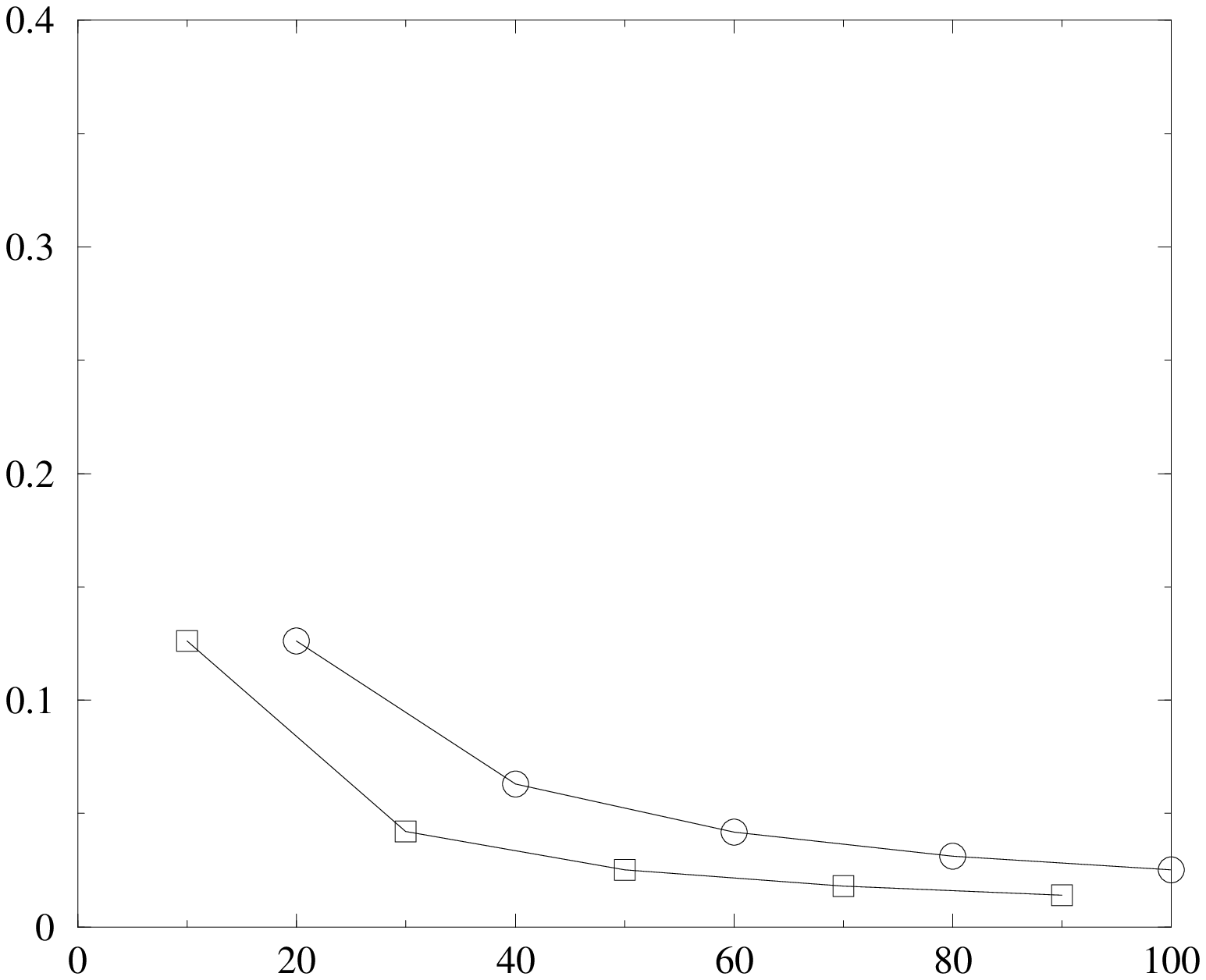}}
\put(165,15){\includegraphics[height=5.0cm,width=4.5cm]{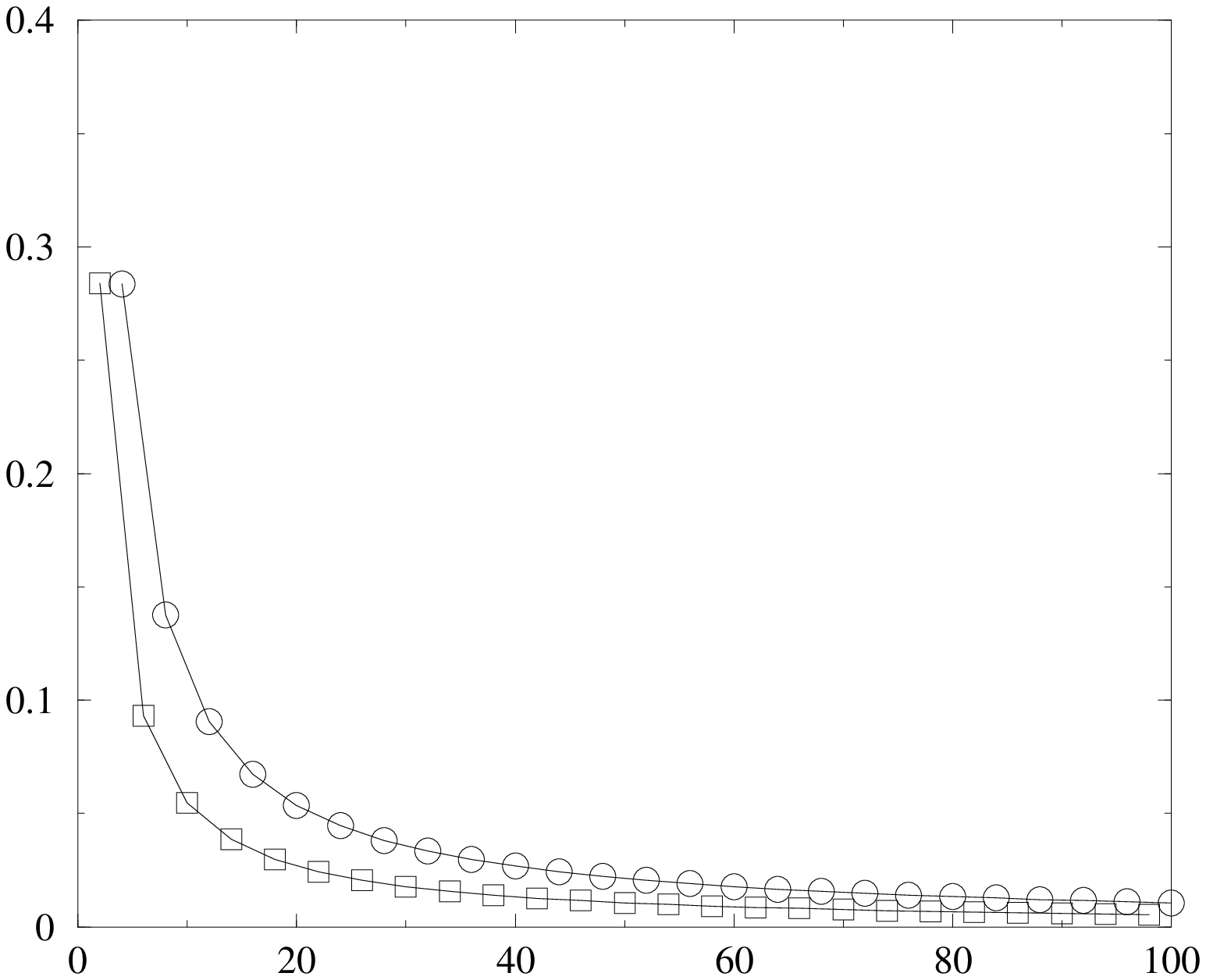}}
\put(315,15){\includegraphics[height=5.0cm,width=4.5cm]{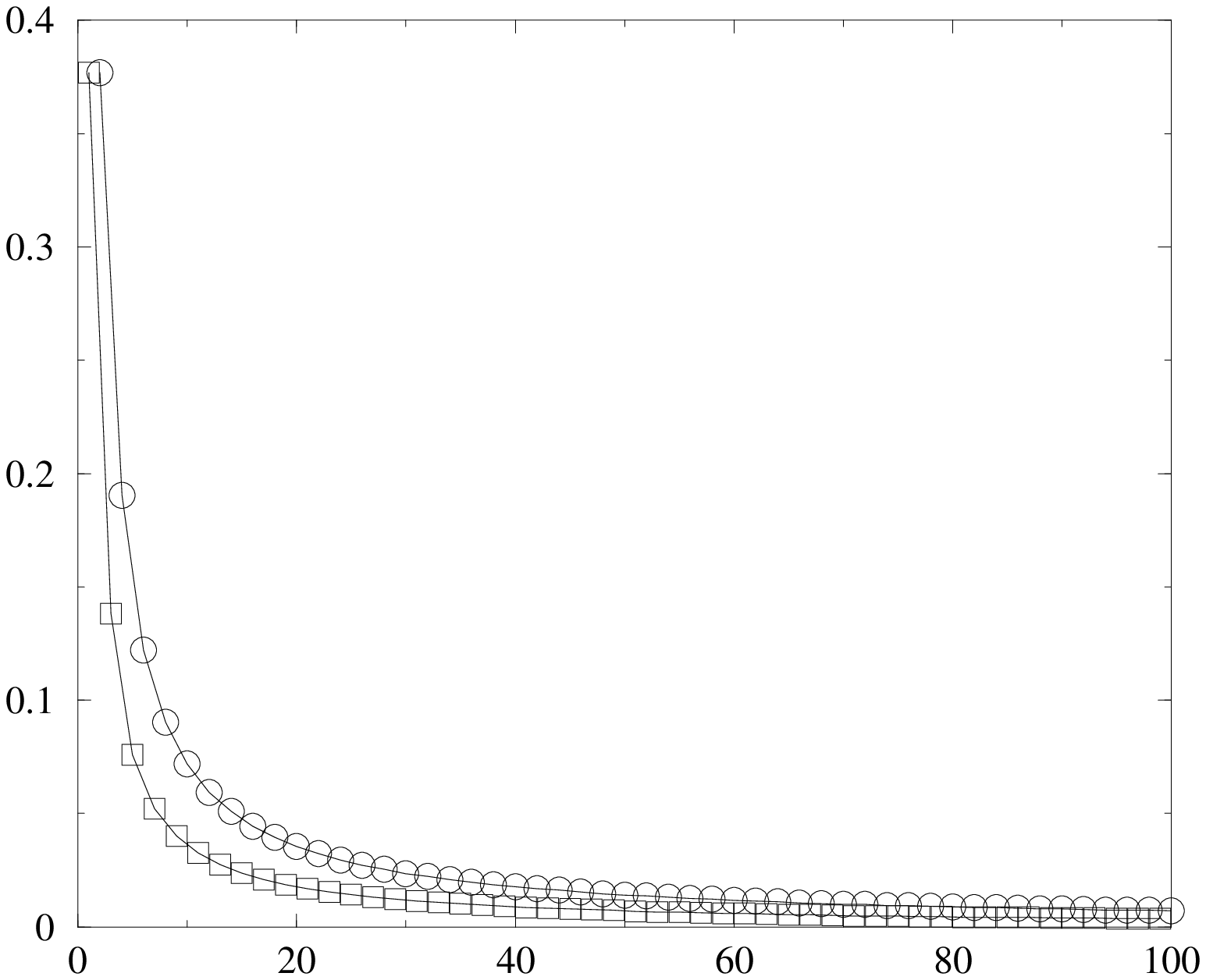}}
\put(80,0){{\small $c$}}
\put(0,90){{\small $\psi$}}
\put(230,0){{\small $c$}}
\put(150,90){{\small $\psi$}}
\put(380,0){{\small $c$}}
\put(300,90){{\small $\psi$}}
\end{picture}
\caption{Fraction of spins with zero field $\psi$ (\ref{eq:psi_def}) for
$\alpha=0.1,0.5,1$ (from left to right). In all figures $T=2$ and $n=2$.
Distinction has been made between even $p$ (circles) and odd $p$ (squares).}
\label{frac0}
\end{figure}

For large $c$, the quantity $\psi$ is expected to vanish. 
This is indeed supported 
by the results of figure \ref{frac0}, where
$\psi$ is plotted as a function of $c$ for $T=2$, $n=2$ and the
memory loads $\alpha=0.1$, $\alpha=0.5$ and $\alpha=1$
respectively. Clearly, for small values of $c$ the quantity $\psi$ can be
rather large indeed, especially in the paramagnetic phase.
It is also clear that the statistics for $\psi$ differ
in the case of an odd versus an even number of patterns $p$, which
we have plotted separately in the figures: for an odd number of
patterns, there can be no zero-valued interactions
$\sum_\mu\xi_i^\mu\xi_j^\mu$, whereas for even $p$
this can be the case; hence the smaller values of $\psi$ for odd
$p$. The magnitude of the difference in $\phi$ at $T=2$ between
finite connectivity and infinite connectivity in figure \ref{frac1}
appears to display a slight nonmonotonicity in $\alpha$ at first
sight, but is consistent with the corresponding values for $\psi$
in figure \ref{frac0}. Here the cases $\alpha=0.1$ and $\alpha=0.5$
correspond to odd $p$, whereas for $\alpha=1$ $p$ is even.

\section{Simulations}
To perform simulation experiments one needs to construct an
explicit dynamical process for the transition probabilities
$W[\bc';\bc]$ between two graph configurations. To this end, one
can easily set up Glauber-type probabilities $W[F_{ij}\bc,\bc]$
that automatically obey detailed balance and which are functions of
the energetic  difference $H_{\rm s}(F_{ij}\bc)-H_{\rm s}(\bc)$,
with $F_{ij}$ a `switch' operator such that $F_{ij}f(c_{ij})=f(1-c_{ij})$
and $F_{ij}f(c_{k\ell})=f(c_{k\ell})$ if $(i,j)\neq (k,\ell)$. Explicit
expressions on such processes can be derived equivalently
to \cite{WSC}. The result is
\begin{eqnarray}
\hspace*{-20mm}
W[F_{ij}\bc; \bc] = \nonumber \\
\hspace*{-5mm}
\frac{1}{2}\left\{ 1- \tanh\left[
\frac{(2c_{ij}-1)}{2} \log \left(\frac{c}{N}\right) - \frac{n}{2}
\log\left(\left\bra {\rm e}^{-\frac{\beta}{c} \sum_\mu \xi^\mu_i \xi^\mu_j
(2c_{ij}-1)\sigma_i \sigma_j }\right\ket \right) \right] \right\}
\label{eq:glaubcij}
\end{eqnarray}
The angular brackets denote a thermal average over the distribution
$p_{\rm f}(\bsigma)\sim\exp[-\beta H_{\rm f}(\bsigma,\bc)]$ while spins are in equilibrium at the
timescale of the graph dynamics. Note that this transition
probability reduces to the one found in \cite{WSC} for $c \to
\infty$.

The practical restrictions for `coupled' types of
simulations, where two nested dynamical processes occur, is that
the total equilibration time can in reality be extremely large. So,
for all practical purposes one is confined to small system sizes,
which, unfortunately induce, in turn, large finite-size effects.
Therefore, one has to accept that experiments will always suffer
from a non-vanishing statistical error and can at most be
satisfactory.

\begin{figure}[t]
\begin{picture}(80,150)
\put(150,10){\includegraphics[height=4.5cm,width=5.0cm]{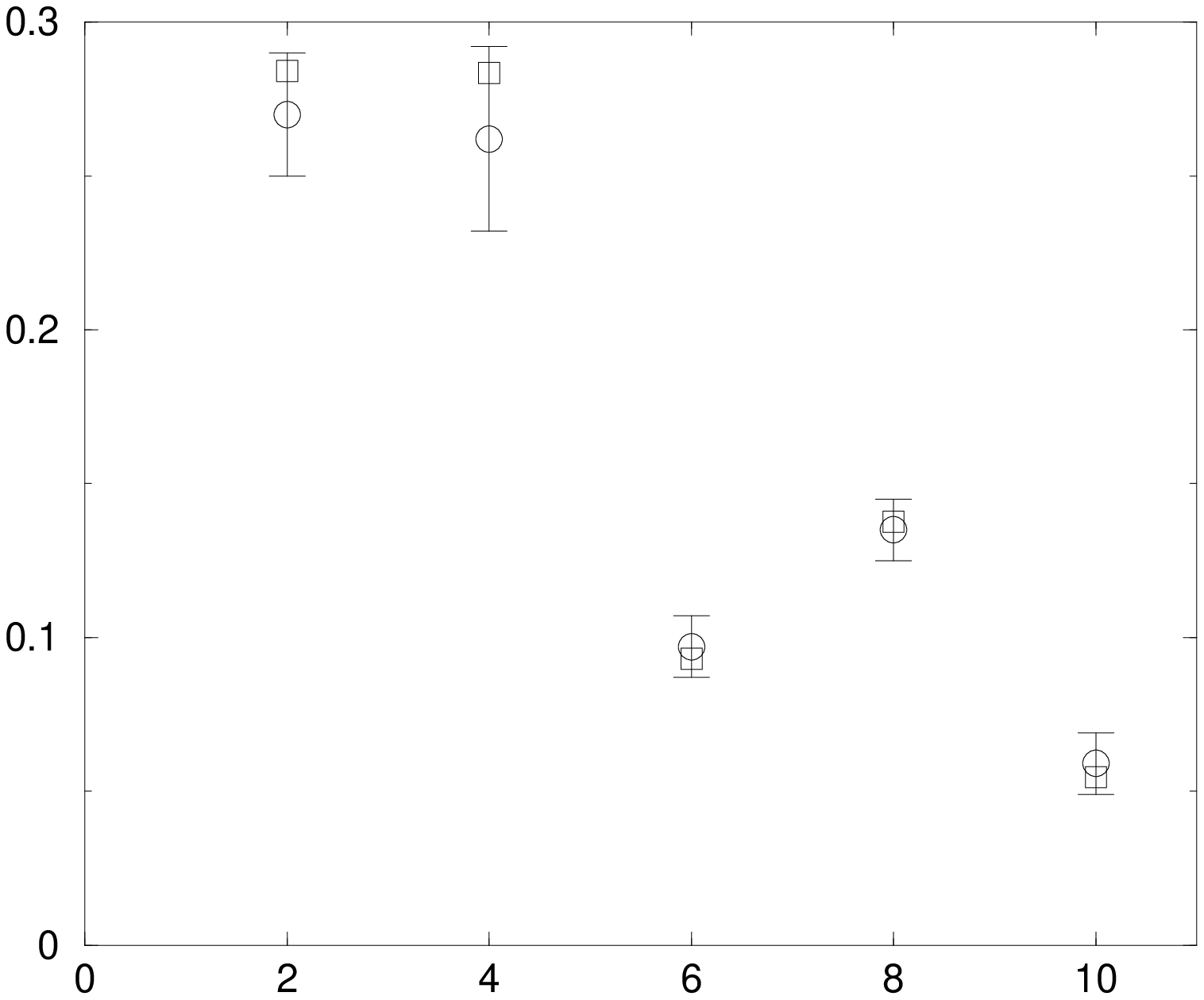}}
\put(130,70){\small $\psi$}
\put(220,0){\small $c$}
\end{picture}
\caption{Theory versus simulations: We plot the fraction $\psi$ of spins with zero local field for $\alpha=0.5$, $T=2$ and $n=2$
(theory: squares, simulations: circles plus error bars) .
Simulations are performed for a system of $N=200$ spins using the Glauber
dynamics (\ref{eq:glaubcij}).}
\label{fig:sims}
\end{figure}

With these practical limitations in mind, we will resort to compare
our theory for the fraction $\psi$ of zero local fields in the
paramagnetic regime for $n=2$ and $T=2$, and for small $c$. In this way
we exploit the fact that in the paramagnetic regime
equilibration times are relatively short. In figure \ref{fig:sims}
we have plotted $\psi$ for the 5 first possible values of $c$ at
$\alpha=0.5$, with $T=2$ and $n=2$. Even though the system size is
still rather modest (we used $N=200$), the results clearly support
the theory.

\section{Conclusions}
It is a relatively new insight that a
wide range of physical complex systems (viewed as random graphs) evolve towards
specific structures and are, surprisingly,  characterised by universal features (e.g.
scale-free degree distributions). Preferential attachment models describing
network growth,
such as the Barab\'asi-Albert algorithm \cite{BarAlrev}, provide possible
explanations for the occurrence of certain network-architectures. 
In the present paper we have 
presented a simple solvable model describing evolving random graphs of a 
fixed number of nodes, with 
finite connectivity, in
which the analytic hurdles imposed by the explicit dynamics are overcome by taking the
adiabatic limit: 
we consider connectivities which evolve much slower than nodes. This
allows us to study directly thermodynamic properties of graphs at equilibrium.

In particular, we have focused our attention on Hopfield neural
network models on random graphs. By a suitable choice of chemical
potential, the average connectivity per neuron is forced to a
finite regime, as opposed to a previous study \cite{WSC} in which
it was infinite. The resulting theory is, traditionally (see
\cite{CoolenSherrington,PCS1,JBC,JABCP,CoolenUezu,FeldmanDotsenko,DotsFranzMezard}),
a finite dimensional replica theory, in which the replica dimension
represents the ratio of temperatures between the fast and the slow
system.

The consequences of the finite connectivity regime, as compared to
the regime of so called extreme dilution, are drastic at the
mathematical level, since a finite set of order parameters is
traded for an infinite dimensional order parameter function. The
case of integer replica dimension, however, is an exception to the
latter statement, simplifying the mathematics, so that
observables can be calculated accurately. In this paper we have
presented phase diagrams and observables for both integer replica
dimension and non-integer replica dimension. For integer replica
dimension, we were able to calculate the fraction of
misaligned spins, as well as the fraction of vanishing fields. The
latter quantity, which vanishes in the case of infinite
connectivity, is the most important discriminator between the
parameter regimes of finite connectivity and extreme dilution.
Qualitatively, the frustration effects are minimised as the
temperature of the slow system decreases, enhancing the retrieval
state of a condensed pattern. Moreover, for larger $n$, phase
transition lines are mainly first order, and the retrieval phase is
dramatically enlarged in comparison with the case $n=0$. For any
nonzero $n$ there is no critical storage capacity beyond which
retrieval is not possible. In other words, as soon as the
statistics for the graph realisations becomes nonuniform, the graph
in principle can organise itself in favour of the retrieval of a
pattern, regardless of the number of patterns present in the
Hopfield interactions (note that this statement is in principle
true for an arbitrarily large but finite $p$, due to the scaling
regime of finite connectivity). 
Our theoretical findings are
supported by simulation results, which, due to the difficulties of
nested equilibrations, we have restricted to a simple
(paramagnetic) region in the phase diagram. All results obtained in
this paper are replica symmetric approximations. However, as in the
case of infinite connectivity \cite{WSC}, we expect replica
symmetry breaking to occur only at sufficiently low values of $n$
i.e. below $n=1$, and small values of $T$ or large values of
$\alpha$.

Several extensions of this work are possible: in a similar
fashion one may also consider evolution models of random poissonian
graphs to e.g.\@  scale-free ones. In the context of the present paper this
amounts to appropriate modifications of our chemical potential
while the remaining theory would remain largely the same.

Furthermore, one may investigate the local stability of non-condensed
pattern retrieval
states once the $c_{ij}$ are equilibrated in favour of one particular pattern,
as was done in \cite{WSC}. To that end, one needs to overcome the problem of
taking the replica limit $n \to 0$ in only one of two spin systems 
$\{ \bsigma_i \}$ and $\{ \btau_i \}$, described by a coupled order function
$P_{\bxi}(\bsigma, \btau)$. So far, we have not been able to solve 
this problem.

\section*{Acknowledgment}
The authors are indebted to A.C.C.\@ Coolen, J.P.L.\@ Hatchett, T.\@ Nikoletopoulos
and I.\@ P\'erez-Castillo for illuminating discussions on finite connectivity
issues.
BW acknowledges financial support from Stichting FOM
(Fundamenteel Onderzoek der Materie) in the Netherlands and NS from
the Ministerio de Educaci\'on, Cultura y Deporte (Spain, grant SB2002-0107)
and the ESF SPHINX program.

\end{document}